\documentclass[iop,apj,letterpaper]{emulateapj}
\pdfoutput=1
\usepackage[left=1in,top=1in,bottom=1in,right=1in]{geometry}
\usepackage{natbib}
\usepackage{url}
\usepackage{amsmath, mathtools}
\usepackage[normalem]{ulem}
\usepackage{color}
\newcommand       \be		{\begin{equation}}
\newcommand       \ee		{\end{equation}}
\newcommand       \bea          {\begin{eqnarray}}
\newcommand       \eea          {\end{eqnarray}}

\newcommand       \kpc		{\,{\rm kpc \,}}
\newcommand       \pc		{\,{\rm pc \,}}

\newcommand       \yr		{\,{\rm yr \,}}
\newcommand       \Myr		{\,{\rm Myr \,}}
\newcommand       \cm		{\,{\rm cm \,}}
\newcommand       \s		{\,{\rm s \,}}
\newcommand       \g		{\,{\rm g \,}}
\newcommand       \kms          {\,{\rm km \,\, s}^{-1}}
\newcommand       \eff          {\epsilon_{\rm ff}}
\newcommand       \etaeff       {\eta}
\newcommand       \tff          {\tau_{\rm ff}}
\newcommand       \tdyn         {\tau_{\rm dyn}}
\newcommand       \tDyn         {\tau_{\rm DYN}}

\newcommand       \R           {r_0}

\newcommand       \trho 	{{\tau_{\rho}}}
\newcommand       \tvr  	{{\tau_{u_r}}}
\newcommand       \tvt  	{{\tau_{v_T}}}

\begin{document}

\title{Star formation in self-gravitating turbulent fluids}
\author{Norman Murray\altaffilmark{1,2} \& Philip Chang\altaffilmark{3},
  } 
\altaffiltext{1}{Canadian Institute for Theoretical Astrophysics, 60
  St.George Street, University of Toronto, Toronto ON M5S 3H8, Canada;
  murray@cita.utoronto.ca} 
\altaffiltext{2}{Canada Research Chair in Astrophysics}
\altaffiltext{3}{Department of Physics, University of
  Wisconsin-Milwaukee, 1900 E. Kenwood Blvd., Milwauke, WI 53211, USA;
  chang65@uwm.edu} 

\begin{abstract}
We present a model of star formation in self-gravitating turbulent
gas. We treat the turbulent velocity $v_T$ as a dynamical variable,
and assume that it is adiabatically heated by the collapse. The theory
predicts the run of density, infall velocity, and turbulent velocity,
and the rate of star formation in compact massive gas clouds. The
turbulent pressure is dynamically important at all radii, a result of
the adiabatic heating. The system evolves toward a coherent spatial
structure with a fixed run of density, $\rho(r,t)\to\rho(r)$; mass
flows through this structure onto the central star or star cluster.
We define the sphere of influence of the accreted matter by
$m_*=M_g(r_*)$, where $m_*$ is the stellar plus disk mass in the
nascent star cluster and $M_g(r)$ is the gas mass inside radius
$r$. The density is given by a broken power law with a slope $-1.5$
inside $r_*$ and $\sim -1.6$ to $-1.8$ outside $r_*$. Both $v_T$ and
the infall velocity $|u_r|$ {\em decrease} with decreasing $r$ for
$r>r_*$; $v_T(r)\sim r^p$, the size-linewidth relation, with
$p\approx0.2-0.3$, explaining the observation that Larson's Law is
altered in massive star forming regions. The infall velocity is
generally smaller than the turbulent velocity at $r>r_*$. For $r<r_*$,
the infall and turbulent velocities are again similar, and both {\em
  increase} with decreasing $r$ as $r^{-1/2}$, with a magnitude about
half of the free-fall velocity. The accreted (stellar) mass grows
super-linearly with time, $\dot M_*=\phi M_{\rm cl}(t/\tff)^2$, with
$\phi$ a dimensionless number somewhat less than unity, $M_{\rm cl}$
the clump mass and $\tff$ the free-fall time of the clump. We suggest
that small values of $p$ can be used as a tracer of convergent
collapsing flows.
\end{abstract}

\section{Introduction}
\label{sec:intro}

The importance of massive stars in a human context became apparent 
when \citet{1957RvMP...29..547B} showed that such stars were
responsible for the production and distribution of most of the heavy
elements that composed the Earth, and which form the building blocks
of life. 

More recently it has emerged that massive stars are important
on a less parochial scale. Star formation on galactic scales is
observed to be slow, in the sense that the time to deplete the supply
of gas from which stars are made is some fifty times the dynamical
time. Quantitatively,
\be \label{eqn: Schmidt Kennicutt} 
\dot M_*=\eta {M_g\over \tDyn},
\ee  
where $\eta\approx0.017$ is a dimensionless constant, and
$\tDyn=R/v_c$ is the dynamical time of a galactic disk of radius $R$
with circular velocity $v_c$
\citep{1998ApJ...498..541K,2008AJ....136.2782L}. Massive stars provide
the feedback that limits the rate of star formation in galaxies. The
feedback comes in the form of radiation pressure, gas pressures in
ionized regions, blast waves from supernovae, and cosmic rays. The
feedback also drives gas out of galaxies, setting the stellar
mass-halo mass relationship for halo masses
$M_h\lesssim10^{13}M_\odot$, e.g.,
\citet{2013arXiv1311.2073H,2013arXiv1311.2910T,2014arXiv1404.2613A,2014MNRAS.442.1545C}.
Understanding how massive stars form is clearly important to our
comprehension of the evolution of the universe, and many of its
components.

Stars more massive than sixty solar masses ($M*>60M_\odot$) live only
$4\Myr$. Since we see stars with masses in excess of $60M_\odot$ outside
their natal dense gas clumps, the early accretion rate onto such a
star must exceed $\approx3\times10^{-5}M_\odot\yr^{-1}$, where we have
assumed that the most massive stars are visible for $\gtrsim1\Myr$, hence
they can accrete for no more than $3\Myr$. In the Milky Way, these
massive stars are usually found in compact $1-10\pc$ radius clusters
containing up to $10^5M_\odot$ of stars.  The accretion rates for
clusters must exceed a few hundredths of a solar mass per year, $\dot
M_*\gtrsim 3\times10^{-2}M_\odot\yr^{-1}$. For example, Cygnus OB2
contains $M_*\gtrsim 4-10\times10^4M_\odot$ inside a half light radius
$r_h=6.4\pc$ \citep{2000A&A...360..539K}, yielding $\dot M_*\approx
0.02M_\odot\yr^{-1}$. The free-fall time of the natal cluster is
\bea  
\tff&\equiv&\sqrt{3\pi\over 32G\rho}\nonumber\\
&=&1\left({3\times10^{4}M_\odot\over M_*}\right)^{-1/2}
\left({R_{1/2}\over6.4\pc}\right)^{3/2}
\left({\epsilon\over0.5}\right)\Myr,\nonumber
\eea  
where all quantities are estimated at the half light radius, and
$\epsilon\equiv M_*/M_g$ is the fraction of natal gas in a dense
cluster-forming region that is converted into stars. The star
formation rate per free fall time, the small-scale analog of $\eta$ in
eqn (\ref{eqn: Schmidt Kennicutt}), is
\bea  
\eff&\equiv&{\dot M_*\over \tff M_g}
={\tff\over 3\Myr}\epsilon\nonumber\\
&\approx&0.16
\left({R_{1/2}\over 6.4\pc}\right)^{3/2}
\left({M_{*,1/2}\over 3.5\times10^4M_\odot}\right)^{1/2}
\left({\epsilon\over 0.5}\right)^{1/2}.
\eea  

This estimate of $\eff$ is consistent with many estimates in the
literature, for example, \citet{2005ApJ...635L.173W}, \citet{2010ApJ...724..687L},
\citet{2010ApJ...723.1019H}, and
\citet{2011ApJ...729..133M}, although there is some disagreement, e.g.,
\citep{2012ApJ...745...69K}. 

The fact that $\eff>>\eta$ is one of the reasons that astronomers
believe that stellar feedback is necessary to explain the slow global
rate of star formation; if $\eff=\eta$, no feedback would be
required. A second motivation for feedback is that simulations on the
scale of galaxies, and on the scale of star forming clumps ($r\sim
1-10\pc$) both find very rapid star formation if no feedback is included.

In this paper we are interested in the physics behind rapid star
formation rather than the physics of feedback. The problem arises from
the high accretion rates, discussed above, required to make massive
stars. The early model of \citet{1977ApJ...214..488S} gave an estimate
of the star formation rate. His model assumed that stars formed from
hydrostatic cores supported by thermal gas pressure. The accretion
rate in his model was independent of time, given by $\dot
M_*=m_0c_s^3/G$, where $c_s\approx0.2\kms$ is the sound speed in
molecular gas, and $m_0=0.975$. Thus the maximum accretion rate was
$\dot M_*\approx2\times10^{-6}M_\odot\yr^{-1}$, too small by a factor of
ten to explain the origin of massive stars.

The difficulty with the accretion rate was overcome by
\citet{1992ApJ...396..631M}, who noted that cores in massive star
forming regions had linewidths that exceeded the thermal line width by
factors of several or more. They pointed out that the accretion rate
in Shu's model was limited by the signal speed of the expanding
collapse wave, then took advantage of the larger signal speed
available if they assumed that the pressure in the fluid was given by
\be \label{eqn: turbulent pressure}
P(r,t)=\rho(r,t) \left[c_s^2+ v_T^2(r)\right],
\ee 
where $\rho(r,t)$ is the local density, $c_s$ is the sound speed, and
$v_T(r)$ is the turbulent velocity a distance $r$ from the center of
the collapse. 

The model was further developed by \citet{1997ApJ...476..750M} and
\citet{2003ApJ...585..850M}. These models kept some of the features of
Shu's model, including the assumption of a hydrostatic initial core,
and the static nature of the turbulent velocity in the equation of
state (\ref{eqn: turbulent pressure}). In particular, they assumed
that $v_T$ was not affected by the collapse, and that it was
independent of time. They did not address the question of how a
hydrostatic core was assembled.

Observations in large GMCs, and low mass cores, find 
\be \label{eqn: turbulent} 
v_T(r)=v_T(R)\left({r\over R}\right)^p
\ee  
with $p\approx 0.5$ \citep{1981MNRAS.194..809L, 1988ApJ...329..392M};
as we will discuss further below, in massive star forming regions
$v_T$ behaves differently, with $p\approx0.2$ and much larger
turbulent velocities at a given separation $r$ than in low mass star
forming regions. The turbulent core models either explicitly assume at
turbulent velocity of the form (\ref{eqn: turbulent}), or assumed
implicitly that it follows a power law in radius so that $P(r)\sim
\rho^\gamma(r)$. 

In a companion paper \citep{2014arXiv1406.4148L} (Paper I) we present
three dimensional hydro and MHD simulations which recover the result
$p\approx 0.5$ in the bulk of the gas. However, we also show that in
small regions around density peaks, which we identify as cores,
$p\approx 0.2$.

In this paper we assume that massive stars form in regions of
turbulent flows in which the gas is converging, so that there is never
any time at which the region is in hydrostatic equilibrium. The infall
therefore extends over a much larger region than that envisioned in the
hydrostatic turbulent core models. 

A more profound difference arises from the fact that we model
the turbulent velocity $v_T(r,t)$ as a dynamical variable, changing in
response to the evolution of the system. This requires that we
introduce a third dynamical equation, the energy equation. However,
we are unable to solve the equations that result. This is not
surprising, because numerical solutions of the full set, carried out
in three dimensions, show that the motion becomes turbulent. 

Suitably chastened, we then make two simplifications. First, we assume
spherical symmetry, reducing the problem from three spatial dimensions
to one. Second, we introduce a closure for the energy equation, that
proposed by \citet{2012ApJ...750L..31R}.

We import the notion of the sphere of influence of the star from
galactic dynamics, and define the radius $r_*(t)$ at which the mass of
gas just equals the (time dependent) mass of the star. With our two
assumptions, and the concept of the sphere of influence, we are able
to find self-similar solutions that describe the collapse of a
self-gravitating turbulent gas. We calculate the run of density
$\rho(r,t)$, of infall velocity $u_r(r,t)$, and of the turbulent
velocity $v_T(r,t)$. We find the striking result that for $r<r_*(t)$,
the density approaches a fixed function of radius, independent of
time, i.e., $\rho(r,t)\to \rho_*(r)\sim r^{-3/2}$. For $r>r_*$,
$\rho(r,t)\sim r^{-k_\rho}$, with $k_\rho\approx 1.6-1.8$.

We show that the acceleration due to the pressure gradient always
tracks the acceleration of gravity. For $r<r_*(t)$, the result is that
both the infall velocity and the turbulent velocity increase with
decreasing radius, as $r^{1/2}$. But for $r>r_*(t)$, the pressure
force exceeds that due to gravity. As a result, both $v_T(r,t)$ and
$u_r(r,t)$ increase slowly with increasing radius, $v_T(r,t)\sim r^p$,
with $p\approx 0.2$. This appears to explain the deviations from
Larson's size-linewidth relation seen in massive star forming regions.

It follows from the fact that the density is independent of time for
$r<r_*$, and from $u_r(r,t)\sim\sqrt{GM_*/r}$, that the mass accretion
rate increases linearly with time, so that
\be  
M_*=\phi M_{cl}\left({t-t_*\over \tff}\right)^2,
\ee  
where $M_{cl}$ is the mass of the star forming clump, and $\tff$ is
the free fall time of the clump.

This paper is organized as follows.  In \S \ref{sec: collapse} we
present a spherically symmetric one dimensional model for the collapse
of a turbulently supported gas cloud consisting of the continuity and
momentum equations, and a simple closure for the energy equation. We
describe approximate analytic solutions to these equations, and
comment on the effects of thermal and magnetic support.  In \S
\ref{sec: numerics} we solve the equations numerically. In \S
\ref{sec:discussion} we briefly compare observations of infall in the
massive star forming regions, with special attention given to
G10.6-0.4. We also discuss previous analytic and numerical work.  We
wrap up with our conclusions in the final section.

\section{SPHERICAL GRAVITATIONAL COLLAPSE MODELS}
\label{sec: collapse}
The governing equations are the continuity equation
\be  
{\partial\rho\over\partial t}+{\bf \nabla}\cdot (\rho{\bf u}) =0,
\ee  
the momentum equation
\be  
{\partial \rho{\bf u}\over \partial t}+ \nabla\cdot \rho{\bf u u}
= -{\bf \nabla} P + \rho{\bf g},
\ee  
and the energy equation
%
\bea \label{eqn: energy} 
&{\partial\over \partial t}&\left[\rho\left({1\over2}u^2
+ {\cal E}\right) \right]
+ \nabla\cdot
\left[
 \rho\left(
  {1\over2}u^2 + {\cal E}
 \right) 
 \cdot {\bf u} + {\bf u}P 
\right]\nonumber\\
&&= \rho {\bf g}\cdot{\bf u}.
\eea  
In these equations $\rho$ is the density, ${\bf u}$ is the fluid
velocity, $P$ is the gas pressure, ${\bf g}$ is the acceleration due
to gravity, and ${\cal E}$ is the internal energy of the gas.

In the companion paper (Paper I), we solve these equations via a three
dimensional simulation. In this paper, we simplify our model
to gain some physical insight into the collapse
process. To do so we will simplify the energy equation. We start by
noting that simulations of both supersonic and subsonic turbulence
have generally found that the properties of the turbulent flow are
universal, in the sense that certain relations are found under a wide
variety of conditions. In particular, simulations find velocity power
spectra that have approximate power law forms, e.g., $P_v(k)\sim
k^{-5/3}$ for subsonic tubulence or $P_v(k)\sim k^{-2}$ for supersonic
turbulence. Correspondingly, $v_T(r)\sim r^{1/3}$ or $v_T(r)\sim
r^{1/2}$.

It is tempting to conclude that the energy equation 
can be replaced by a simple closure relation specifying the
turbulent velocity, namely $v_T(r,t)=v_T(r)$, where $v_T(r)$ is given
by Larson's law. This is what has been done in most previous analytic
work on turbulent collapse.

This temptation should be resisted. It is well known that turbulence
decays on a turnover time $r/v_T(r)$, and it is believed that in star
forming galaxies the energy lost in this decay is replaced by stellar
feedback, e.g., from supernovae. One could argue that this leads to a
steady state, i.e., that $v_T(r,t)=v_T(r)$. However,
\citet{2012ApJ...750L..31R} showed that compression of turbulent gas
also drives turbulence. Turbulent gas in a converging flow will be
compressed, which tends to increase $v_T$, while $v_T$ will also decay
due to dissipation.  Therefore, we use a closure scheme suggested by
\citet{2012ApJ...750L..31R} that captures both the decay and the compressive
driving. 

In the absence of collapse, undriven turbulence
decaying on an eddy turnover time is described by
\be  \label{eqn: decay}
{1\over2}{dv_T^2(r)\over dt} = -\eta {v_T^3\over r},
\ee  
with $\eta $ a dimensionless constant of order unity. 

Collapse alters the turbulence. Ignoring for the moment the cascade of
energy from large scales to small that is responsible for
(\ref{eqn: decay}), a reduction in the radius from $R$ to $r$ results in a
scaling of the velocity $v_T(r) = v_T(R)\times (R/r)$, i.e.,
\be  
\left({dv_T\over dt}\right)_{\rm AH}=-v_T{u_r\over r},
\ee  
where $u_r=dr/dt$ is negative for a collapse. Thus the scaling relation suggests
that a collapse will tend to increase the turbulent
velocity. Robertson \& Goldreich refer to this as ``adiabatic
heating'', and show that such heating occurs in cosmological
simulations in which the scale factor decreases with time, mimicking a
gravitational collapse.

Combining this driving term with the turbulent damping,
\be \label{eqn: simple energy} 
{d v_T\over dt} = 
-\left(1+\etaeff {v_T\over u_r}\right){v_Tu_r\over r}.
\ee  
We use this equation to replace the energy Equation (\ref{eqn: energy}).

\subsection{Fundamental Equations}
In Paper I we show that on parsec and smaller scales, the collapse is,
very roughly, spherically symmetric. Motivated by this and our desire
to simplify our model, we employ spherical coordinates. Both
simulations and observations of star forming regions show the
presence of star forming regions that have cylindrical geometry, as
well as other regions with spherical geometry, or a combination of
both. We have made some initial forays into cylindrical geometry, but
for simplicity we stick to spherical geometry in this paper.

The continuity
and momentum equations become
\be \label{eqn: continuity}
{\partial\rho\over \partial t}+{1\over r^2}{\partial\over\partial r}
\left(r^2u\rho\right) = 0,
\ee 
and
\be \label{eqn: momentum}
{\partial u_r\over \partial t}+u_r{\partial u_r\over \partial r}
+{1\over \rho}{\partial \rho (v_T^2(r,t)+c_s^2)\over \partial r}+{GM(r,t)\over r^2}=0,
\ee 
while the energy equation is
\be \label{eqn: 1D energy}
{\partial v_T\over \partial t}+ u_r{\partial v_T\over\partial r} +
\left(1+\etaeff {v_T\over u_r}\right){v_Tu_r\over r}=0.
\ee 

In these equations $u_r(r,t)$ is the mean radial velocity of the
fluid, excluding the turbulent velocity, which is denoted by
$v_T(r,t)$. In the rest of this section only, we will neglect the
sound speed $c_s$ to simplify the presentation. In any case, we find
that its inclusion has little effect on the results.

The mass $M(r,t)=M(0,t)+M_g(r,t)$ where
\be \label{eqn: mass}
M_g(r,t)= 4\pi\int_{0^+}^r r^2\rho(r,t)dr
\ee 
is the gas mass.

We will employ $M_*(t)$ rather than $M(0,t)$, with the
understanding that $M_*$ may refer to the mass of a single star or to
a star cluster. For later use we define the radius or sphere of
influence of the nascent star or star cluster by
\be  \label{eqn: rstar}
M_*(t) = 4\pi\int_{0^+}^{r_*(t)} r^2 \rho(r,t) dr.
\ee  
Thus $r_*(t)$ is the radius at which the enclosed gas mass equals the
mass in stars; this makes sense if the distribution of young stars is more
centrally concentrated than the gas, which is true on large enough
scales. For a pure power law density distribution $\rho(r)\sim r^{-k_\rho}$,
\be 
r_*(t)=R\left[{M_*(t)\over M_{cl}}\right]^{1/(3-k_\rho)},
\ee 
where $R$ is a fiducial radius, and $M_{cl}$ is the mass of gas
inside that radius. We have in mind the thickness of the ribbon-like
features found in hydrodynamic simulations, e.g., those in paper I,
for $R$.


\subsection{Solution}
We will concentrate on a solution from the time when a star first
forms.  We define scaled variables
\be 
x\equiv {r\over r_0}
\ee 
and
\be 
y\equiv{t\over\tDyn},
\ee 
where $r_0$ is a fiducial radius, and $\tDyn=R/u_r(R)$. 

We define
\be  
p\equiv{r\over v_T}{\partial v_T\over \partial r},
\ p'\equiv{r\over u_r}{\partial u_r\over \partial r},\ {\rm and}
\ k_\rho\equiv-{r\over \rho}{\partial \rho\over \partial r},
\ee  
anticipating that each will vary only slowly with radius. In other
words, we look for solutions in which the density, infall velocity,
and turbulent velocity are all approximately power laws in both $r$
and $t$,
\be 
\rho(r,t)=\rho_0x^{-k_\rho} y^\alpha,
\ee 
%
\be 
u_r(r,t)=u_0 x^{p'} y^\beta,
\ee 
and
\be 
v_T(r,t)=v_0 x^p y^\gamma.
\ee 
where we have defined $\rho_0\equiv\rho(r_0,\tDyn)$, with similar
expressions for $u_0$ and $v_0$.

For an infall solution, we expect that the magnitude of the infall
velocity increases with time; mass accumulates in the central object,
which then exerts a gravitational force on the surrounding material
that increases with time. Our energy closure then ensures that the
turbulent velocity will increase (at a fixed radius) with increasing
time, so that
\be 
\beta>0,\quad\gamma>0.
\ee 

We define $\tilde\rho(t)\equiv\rho_0 y^\alpha$, so that
$\rho(r,t)=\tilde\rho(t) x^{-k_\rho}$. 

\subsection{Solution at Small Radius, $r<r_*(t)$} \label{sec: inside}
The continuity equation becomes
\be  \label{eqn: continuity below rstar}
{\partial\tilde\rho\over\partial t} x^{-k_\rho}+{u_0\rho_0\over
  r_0}x^{p'-k_\rho-1}y^{\alpha+\beta}
\left[2+p'-k_\rho\right]=0.
\ee  
As $r\to0$, the second term grows more rapidly than the first term,
since $p'<1$; in fact we will show that $p'=-1/2$. Thus at small radii
($r<<r_*)$ satisfying this equation demands that both terms go to zero
independently.  For the larger, second term, this implies
\be \label{eqn: rho u}
k_\rho=2+p'.
\ee 
Similarly, for the first term, we find 
\be  \label{eqn: density attractor}
{\partial \rho(r,t)\over \partial t}=0.
\ee  
This is a very striking result: self-similar gravitational collapse
(in which the density and infall velocity follow power laws in radius)
evolves to an attractor, in which the density does not vary with
time. Note that the argument has made no reference to the type of
pressure support, i.e., thermal, turbulent, or magnetic.

\subsubsection{Using the energy closure to relate $u_r$ to $v_T$}\label{sec:energy closure}
Using the power law ansatz, Equation (\ref{eqn: 1D energy}) becomes
%
\bea  \label{eqn: coupling}
&\gamma&{v_0\over\tDyn}y^{\gamma-1}x^p+{u_0v_0\over r_0} 
y^{\beta+\gamma}x^{p+p'-1}
\times\nonumber\\
&&\quad\left[p+1+ \left({v_T(r,t)\over u_r(r,t)}\right)\eta\right]=0.
\eea  

We can then show
\be \label{eqn: eliminate} 
v_T(r,t)=-u_r(r,t){1+p\over \eta}-{\gamma\over\etaeff}{r\over t}.
\ee  

At small $r$ and large $t$, the result (\ref{eqn: eliminate}) implies
that $v_T \sim u_r$, so that (\ref{eqn: eliminate}) is the analog of
the result in \citet{2012ApJ...750L..31R} that $v_T\sim
a^{\tilde\beta-1}$, where $a$ is the scale factor in their
cosmological simulations and $\tilde\beta$ is given by their Equation
(9). We note that this result implies
\be \label{eqn: energy exponent}
p=p',\quad \beta=\gamma, \quad {\rm for}\ r<r_*.
\ee 

\subsubsection{The Momentum Equation at Small Radii}
We start by noting that
\be 
M(r,t)=M_*(t) + {4\pi\over 3-k_\rho}r_0^3\rho_0x^{3-k_\rho},
\ee 
where the second term on the right hand side does not depend on time,
a result that follows from equation (\ref{eqn: density attractor}).

The momentum equation is
%
\bea  \label{eqn: puzzle}
&&\left(
{u_0\over \tDyn}y^{\beta-1}x^{p'}+{4\pi G\rho_0r_0\over 3-k_\rho}x^{1-k_\rho}
\right) +\nonumber\\
&&\quad\Bigg [
{u_0^2\over r_0}p'y^{2\beta}x^{2p'-1}+{v_0^2\over
  r_0}(2p-k_\rho)y^{2\gamma}x^{2p-1}+\nonumber\\
&&\qquad{GM_*(t)\over r_0^2}x^{-2}\Bigg]=0.
\eea  
We have written equation (\ref{eqn: puzzle}) so as to group together
terms that scale with $r$ in the same manner. From equations
(\ref{eqn: continuity below rstar}) and (\ref{eqn: energy exponent}),
$1-k_\rho=-(1+p)$; since we expect gravity to cause the infall to
increase the magnitude of $u_r$ as $r$ decreases, $p'=p<0$. It follows
that $r^{1-k_\rho}=r^{-1-p}< r^{2p-1}$, i.e., the pressure gradient term is
larger in magnitude than the gas self-gravity term, and also larger
than $|\partial u_r/\partial t|$. The pressure gradient
term and the advective term must together balance the stellar
gravitational term, so we group those three terms together in the
square brackets. On setting the sum of the three terms to zero, we
have
\be \label{eqn: u_r small r}
u_r=-\Gamma
\sqrt{GM_*(t)\over r},
\ee 
where we have used Equation (\ref{eqn: eliminate}), and defined
\be \label{eqn: Gamma}
\Gamma\equiv\left[(k_\rho-2p)\left({1+p\over\etaeff}\right)^2-p'\right]^{-1/2}.
\ee 

Combining Equation (\ref{eqn: u_r small r})
with Equation (\ref{eqn: rho u}) shows that 
\be \label{eqn: vanishing r predictions} 
p'=p=-{1\over2},\quad k_\rho={3/2},\quad (r<r_*).
\ee 
Equation (\ref{eqn: u_r small r}) also shows that $M_*(t)\sim
t^{2\beta}$.

The remaining two terms, the
time rate of change of $u_r$ and the gas self-gravity, inside the
parentheses in equation (\ref{eqn: puzzle}) must separately
cancel, or
\be 
{\partial u_r\over \partial t}=-{4\pi G \rho(r_0)r_0\over 3-k_\rho}x^{1-k_\rho}.
\ee 
This shows that the infall velocity increases linearly with time
($\beta=1$) after
the central object forms. Integrating,
\be \label{eqn: second velocity}
u_r(r,t)=u_r(r,0)-{4\pi G \rho(r_0)r_0\over 3-k_\rho}x^{-1/2}t.
\ee 

The two expression (\ref{eqn: u_r small r}) and (\ref{eqn: second
  velocity}) must agree at late times, so that:
\be 
M_*(t)=\Gamma^{-2}
\left({4\pi\rho(r_0)\over 3-k_\rho}\right)^2Gr_0^3 t^2.
\ee 
We show in the next section that $k_\rho\approx1.6-1.7$ for $r>r_*$.
This allows us to connect the density at any $r<r_*$ to the density
$\rho(R)$ at $r=R$, 
\bea  
\rho(r)&=&\rho(R)\left({r\over r_*}\right)^{-3/2}
\left({r_*\over R}\right)^{-k_\rho}\nonumber\\
&=&\rho(R)\left({r\over R}\right)^{-3/2}\cdot
\left({r_*\over R}\right)^{3/2-k_\rho}
\eea  
The fact that $3/2-k_\rho<<1$ suggests that
we define 
\be  
\psi(k_\rho, r_*(t)) \equiv \left({r_*\over R}\right)^{3/2-k_\rho}.
\ee  
Because $k_p-3/2\approx 0.1$, the dependence of
$\psi$ on $r_*$ is very weak; if $r_*=10^{-5}R$, $\psi=3$.

Using this, the stellar mass at time $t$ can be written in terms of
the clump mass as
\be \label{eqn: mass vs time}
M_*(t)=\psi\cdot\,\left({\pi\over 2\Gamma}\right)^2
M_{cl}\left({t-t_*\over\tff}\right)^2,
\ee 
where we now explicitly display $t_*$, the time at which the central density diverges, or that
at which the central star forms.

It follows that
\be \label{eqn: r_star}
r_*(t)=\psi^{2/3}\cdot\,\left[{\pi\over2\Gamma}\right]^{4/3}
\left({t-t_*\over\tff}\right)^{4/3}R,
\ee 
where to a first approximation it is acceptable to treat $\psi$ as a
constant, independent of both $r_*$ and time.

It is instructive to compare the acceleration terms in the momentum
equation, normalizing to $g\approx G M_*(t)/r^2$.  The acceleration
due to the pressure gradient is
\be  \label{eqn: small r pressure}
{1\over g}\left({1\over\rho}{\partial P\over \partial r}\right)
\approx(2p-k_\rho)\left({v_T\over u_r}\right)^2\Gamma^2
\approx-{5\over 5+4\etaeff^2},
\ee  
while the net acceleration of the gas is approximately
\be  
{1\over g}{du_r\over dt}\approx p'\Gamma^2\approx-{4\etaeff^2\over 5+4\etaeff^2}.
\ee  
For $\etaeff = 2/3$, the ratio between the acceleration due to the
pressure gradient and gravity is $(\partial P/\partial r)/\rho g
\approx 0.74$ (equation [\ref{eqn: small r pressure}]), i.e., they nearly
balance each other.  Hence, the net acceleration of the gas is only
roughly one quarter that of gravity. In other words, for $r<r_*$,
gravity dominates the dynamics, but it is nearly balanced by turbulent
pressure, and the infall velocity is substantially smaller than the
free fall velocity.

In the next section we show that for $r>r_*$, both $p'$ and $p$ are greater
than zero; the turbulent pressure decelerates the infall, and the
turbulent velocity tracks the infall velocity, as prescribed by
Equation (\ref{eqn: simple energy}). As the flow crosses $r_*$, the
acceleration due to gravity, which we will show scales as $\sim r^{-0.6}$
for $r>r_*$, transitions to scaling as $r^{-2}$ for $r<r_*$.  The
infall velocity responds to this increased gravity rapidly, i.e. in a
moderately small fraction of the local dynamical time; the turbulent
velocity follows, but with a substantial lag (of order the local
dynamical time), eventually reaching $p=p'$, as enforced by our energy
closure, Equation(\ref{eqn: simple energy}). The acceleration due to the
pressure gradient tracks that due to gravity at all radii. However for
$r>r_*$ the pressure term dominates the gravity term, while for
$r<r_*$, the gravity term is roughly twice the pressure term, as we
have just seen.



\subsubsection{Summary at $r<<R$}

Gathering the results, we have shown that the turbulent velocity
tracks the infall velocity, that $k_\rho=2+p'$ (from the continuity
equation), and $p'=-1/2$ for $r<r_*(t)$; it follows that $k_\rho=-3/2$ for
$r<r_*(t)$.

The mass accretion rate for $r<r_*(t)$ is roughly independent
of radius,
\bea  \label{eqn: flat mdot}
{dM_g(r,t)\over dt}&=&4\pi r^2 \rho(r,t)u_r(r,t)\nonumber\\
&\approx& 4\pi \R^2\rho_f(\R)u_r(\R,t),
\eea  
where we used the fact that $2-k_\rho+p'=0$ for $r<r_*$.  

Although the density $\rho(r,t)=\rho(r)$, i.e., it is independent of
time, the accretion rate does vary with time since the infall velocity
$u_r(t,r)$ does. In other words, the accretion rate does not vary with
radius (for $r<r_*$) at a fixed time, but it does increase linearly
with time at a fixed radius.

\subsection{Solution for $r_*<r\lesssim R$}
It is useful to define times scales for the density, radial velocity,
and turbulent velocity to change by a factor of order unity:
\be  
\tau_\rho\equiv\left({1\over \rho}{\partial \rho\over \partial t}\right)^{-1},
\ee  
\be  \label{eqn: tau u_r}
\tau_{u_r}\equiv\left({1\over u_r}{\partial u_r\over \partial t}\right)^{-1},
\ee  
and
\be  
\tau_{v_T}\equiv\left({1\over v_T}{\partial v_T\over \partial t}\right)^{-1}.
\ee  
Well away from $r=r_*$ and $r=R$, we expect that these should be of
order the local dynamical time
\be  
\tdyn\equiv{r\over v_r}.
\ee  

The momentum equation can be written as
\be  
{\tdyn\over\tau_{u_r}}+p'+\left({v_T\over u_r}\right)^2(2p-k_\rho)
+\left({v_T\over u_r}\right)^2{GM/r\over v_T^2}=0.
\ee  
Since the turbulent velocity tracks the infall velocity, $v_T/u_r$
varies slowly with $r$; for $r_*<< r<<R$, we expect $\tdyn/\tau_{u_r}$
to vary slowly with $r$ as well. Then
\be  \label{eqn: large r velocity}
v_T(r,t)=\Gamma'\sqrt{GM(r,t)\over r},
\ee  
where 
\be  \label{eqn: Gprime}
\Gamma'\equiv \left|{v_T\over u_r}\right|\Gamma
=\left[
k_\rho-2p-
\left(p'+{\tdyn\over \tvr}\right)
\left({u_r\over v_T}\right)^2
\right]^{-1/2}
\ee  
is a slowly varying function of $r$ for $r>r_*$. Note that
$\Gamma^{\prime 2}\approx \alpha_{\rm vir}$, the virial parameter.

Because $\Gamma'$ is a slowly varying function of $r$, it follows from
Equation (\ref{eqn: large r velocity}) that 
\be  \label{eqn: large r momentum} 
2p=2-k_\rho, \quad r>r_*.
\ee  

The continuity equation can be written as
\be  
{\tdyn\over \tau_\rho}+2+p'-k_\rho=0.
\ee  
Combining this with Equation (\ref{eqn: large r momentum}), we
find
\be \label{eqn: large r p prime} 
p'=\left|{\tdyn\over\trho}\right| - 2p, \quad r>r_*.
\ee  

In the appendix we argue that if the outer boundary condition on the
density is $\rho(R,t)=\bar\rho$, we can approximate
\be  \label{eqn: trho}
\left|{\tdyn\over\trho}\right|\approx
\left|{\ln r/r_*\over\ln R/r_*}-1\right|.
\ee  

In the Appendix we show that 
\be  
p(r) = {1\over2}k_\rho -{1\over2}\left({u_r\over  v_T}\right)^2
\left[{\tdyn\over \tvr}+p'+\Gamma^{-2}(0)\right],
\ee  
and estimate $p\approx0.1-0.2$.

From Equation (\ref{eqn: large r momentum}), we find
\be  
k_\rho \approx1.6-1.8,
\ee  
again depending on the initial conditions and on $\etaeff$.

Unlike the case at $r<r_*$, here the dynamics are controlled by the
pressure gradient, and not by gravity. The ratio
\be \label{eqn: pressure gradient ratio} 
{1\over g}\left(
{1\over\rho}{\partial P\over \partial r}
\right)=-(k_\rho-2p)\Gamma'^2\approx-3 \,{\rm to}\, -4 
\ee 
for $v_T\approx1.6 \sqrt{GM(R)/R}$, the fiducial value used in the
numerical simulations below, and $p \approx 0 - 0.2$, motivated by the
results of the Appendix.

The force due to the steep pressure gradient (or more precisely the
density gradient, since $k_\rho >> |2p|$) is larger than the force of
gravity.  This outward-directed force causes the infall velocity to
decrease inward, for $r_*<r<<R$. This is quantified by combining
Equation (\ref{eqn: large r p prime}) with Equation (\ref{eqn: trho}). At
$r_*\lesssim r<<R$, we expect $p'\approx1-2p>0$, so that $|u_r|$
decreases with decreasing radius.

This result, that $|u_r|$ decreases inward for $r>r_*$, is in contrast to
self-similar solutions with an assumed fixed turbulent velocity. In
such theories, the infall velocity is either zero outside the radius
of the expanding collapse wave \citep{1977ApJ...214..488S,2003ApJ...585..850M} or has
a magnitude that {\em increases} inward even at large radii
\citep{2004ApJ...615..813F}.

In the next section, we use direct numerical integration of the 1D
equations to show that $0.1\lesssim p\lesssim0.25$, depending on the
outer boundary conditions and on $\etaeff$, similar to the result of
3D simulations in Paper I.

This value of the turbulent exponent is much smaller than the exponent
in non-self-gravitating supersonic turbulence, $p=1/2$, and smaller
than the Kolmogorov value $p=1/3$ for subsonic turbulence. Recall that
the Kolmogorov value follows from conservation of energy in the
cascade from large scales to small scales. In the case of supersonic
collapse, the exponent is smaller than the value one would derive
under the assumption that the kinetic energy in the cascade is
conserved. Thus, the kinetic energy on small scales is larger than
that produced by an energy conserving cascade. 

The reason is simple: a fraction of the ordered inflow kinetic energy,
and a fraction of the potential energy released in the collapse, is
being converted into turbulent kinetic energy. In our simple theory,
this fraction is set by the closure relation, Equation (\ref{eqn: simple
  energy}), which captures the behavior seen in the numerical
simulations of \citet{2012ApJ...750L..31R}. This extra energy input
boosts the turbulent velocity relative to that in an energy conserving
cascade.

Inside the sphere of influence $r_*$ the conversion of gravitational
potential energy into turbulent energy is even more dramatic, as we
saw above ($p=-1/2$), so that the turbulent velocity {\em increases} with
decreasing $r$.

Simulations of non-self-gravitating turbulence consistently show
$p\approx0.5$, as do observations of GMCs
\citep{1981MNRAS.194..809L,2004ApJ...615L..45H} and low mass star
formation regions \citep{1983ApJ...270..105M, 1992A&A...257..715F} in
the Milky Way. We suggest that measurements of the scaling of
turbulent velocity can be used as a sign post for collapse, a point we
return to below.

As in the case $r<r_*$, we have found that the dynamics converges to a
run of density that varies only on the large scale dynamical time. In
this case (for $r>r_*$) $k_\rho$ is slightly larger than the value
$k_\rho=1.5$ found for $r<r_*$.

Unlike the small $r$ case, however, we also find that the run of
velocity (either $u_r(r)$ or $v_T(r)$) evolves rapidly to a solution
that varies only on the dynamical time evaluated at the outer scale
$R$, i.e., for $r_*<r<R$ the velocity increases only very slowly with
time. This variation, on the global dynamical time, is simply a result of
the change in the mass contained inside $r=R$. For $r_*<<r<<R$ the
turbulent velocity is well described by a power law (with
$p\approx0.2$) while the infall velocity shows substantial deviations
from a simple power law, as given by eqns. (\ref{eqn: large r p
  prime}) and (\ref{eqn: trho}).

It follows that the mass accretion rate for $r>r_*$ varies fairly rapidly
with radius,
\be  \label{eqn: larger r mdot}
{dM_g(r,t)\over dt}=4\pi r^2\rho(R)u_r(R,t)\left({r\over R}\right)^{2-k_\rho+p'},
\ee  
where $2-k_\rho+p'\approx 0.2-0.3$. We have allowed
for the possibility that the infall velocity at $R$ increases with
time, as the growing mass of the clump will tend to accelerate the
infall, but this increase is on the (large scale) dynamical time, and
hence proceeds rather leisurely; essentially, the mass accretion rate
is independent of time for $r>r_*$, but not for $r<r_*$!

%
In a more realistic model, including the effects of thermal gas
pressure, magnetic pressure, and magnetic tension, as well as three
dimensional effects, this mass accretion rate will be reduced. We
discuss these effects briefly in the next subsection.

\subsection{The Effects of Thermal and Magnetic Pressure, Rotation, and
  Magnetic Tension} There are several factors that will reduce the
mass accretion rate in a more realistic model. First, we do not expect
that gas will fall onto the star forming region from every direction,
as we have assumed so far. In the three dimensional simulations in
Paper I, we find that the infall covers a solid angle
$\Omega_{in}<4\pi$ steradians, which we account for by a factor
$f_\Omega\equiv\Omega_{in}/(4\pi)$. In our MHD simulations, the
fraction of the sky (as seen from the central object) over which
accretion occurs is yet smaller, as the magnetic field line tension
inhibits accretion across field lines, described by a factor
$f_{\Omega_B}\le1$. Finally, the pressure associated with the magnetic
field energy density provides an outward pressure gradient much like
the turbulent pressure.

Rotational support will tend to slow the rate of accretion. However,
observations suggest that the rotational velocity $v_c$ in star
forming cores is much smaller than the turbulent velocity, e.g.,
\citet{2011A&A...525A.151B,2014ApJ...785...42P}, who find
$v_c/v_T\lesssim0.3$ at $r\approx 0.05\pc$. This
appears to hold down to scales of order $1000{\rm AU}$ or
$5\times10^{-3}\pc$, below which there is evidence for rotating disks,
e.g., \citet{2000ApJ...537..283V}.

We can include the effects of thermal pressure, and, in an approximate
way, some of the effects of magnetic pressure and tension. All three
tend to reduce the mass accretion rate, although for the massive star
formation regions we consider here, with large virial velocities and
high Mach numbers, the effects of thermal pressure are small.

The effects of magnetic pressure are potentially very significant. We
expect that the strong turbulence generated in the converging flows we
consider will tend to magnify the ambient magnetic field, possibly
reaching equipartition, i.e., a magnetic pressure equal to $\rho
v_T^2$, or a substantial fraction thereof. This effect can be
incorporated in a very rough way by assuming a pressure of the form
\be \label{eqn: magnetic pressure} 
P(r)=\rho(r)\left[c_s^2+\phi_Bv_T^2\right],
\ee 
where $\phi_B=1+B^2/(8\pi\rho v_T^2)$. This expression does not
entirely capture the effect of the magnetic pressure, since it implies
that the magnetic field decays when the turbulent velocity does,
contrary to what we expect. This might be important if the magnetic
pressure becomes large enough to reduce the infall velocity of the
flow; reducing $u_r$ will reduce $v_T$, and hence in (\ref{eqn:
  magnetic pressure}), the pressure associated with $B$. In fact, once
the turbulence generates the magnetic field, the decay time of the
magnetic field need not be related to the decay time of the
turbulence.

We denote the fraction by which the magnetic pressure decreases the
accretion rate by $f_{B^2}$; we show below, using our simple model,
that for a plasma $\beta\approx1$, $f_{B^2}\approx 1/2$.

Accounting for all these effects, the expression for the stellar mass becomes
\bea  
M_*(t)&=&f_\Omega f_{\Omega_B} f_{B^2}\psi\left({\pi\over
  2\Gamma}\right)^2
M_{\rm  cl}\left({t-t_*\over\tff}\right)^2\\
&=&\phi M_{\rm  cl}\left({t-t_*\over\tff}\right)^2.
\eea  

\section{Numerical results}\label{sec: numerics}
We numerically integrate the full set of equations (\ref{eqn:
  continuity}), (\ref{eqn: momentum}), and (\ref{eqn: 1D energy}),
supplemented by (\ref{eqn: mass}) using a simple first order upwind
code.

Our initial conditions are that the density is constant,
$\rho(r,t=0)=\bar\rho$, while the radial and turbulent velocities follow
a power law,
\be  
u_r(r,t=0)=-|u_r(R)|\left({r\over R}\right)^{p\prime}
\ee  
and
\be  
v_T(r,t=0)=v_T(R)\left({r\over R}\right)^p,
\ee  
with fiducial values $\bar\rho=3.0\times10^{-22}\g\s$,
$v_T(R) = 6.0\kms$, $u_r(R)=-3.0\kms$ and $p=p'=1$.  The default value
for the eddy decay time-scale is $\etaeff=2/3$. We have explored
variations around these fiducial values, and report on some of the
results below. The boundary radius $R=3-10\pc$ for most of our
runs. The initial mass is $M(R,0)=4\pi R^3\bar\rho/3\approx
1.8\times10^4(R/10\pc)^3M_\odot$, and the corresponding Keplerian velocity is
\be  
v_k(R,0)=\sqrt{GM(R,0)/R}\approx2.8\kms.
\ee  

Our outer boundary conditions specify the turbulent velocity and
density, $v_T(R,t)=v_T(R)$ and $\rho(R,t)=\bar\rho$. The inflow
velocity $u_r(R,t)$ is set to be a power law extrapolation of
$u_r(R-dr, t)$, where $dr$ is the grid scale. The inner boundary
conditions are $u_r(r_{in},t)=u_r(r_{in}+dr,t)$,
$v_T(r_{in},t)=v_T(r_{in}+dr,t)$, and $(r_{in})^2\rho(r_{in}, t) =
(r_{in}+dr)^2\rho(r_{in}+dr)$, i.e., we insist that the mass flowing
into the innermost shell $\dot M(r,t)$, which is deposited onto the
central star, be the same as the mass flow rate into the surrounding
shell.

We have integrated the equations both with and without including the
effects of a finite sound speed. The results are virtually
identical to those when the sound speed is neglected.

\begin{figure}
\epsscale{1.2}
\plotone{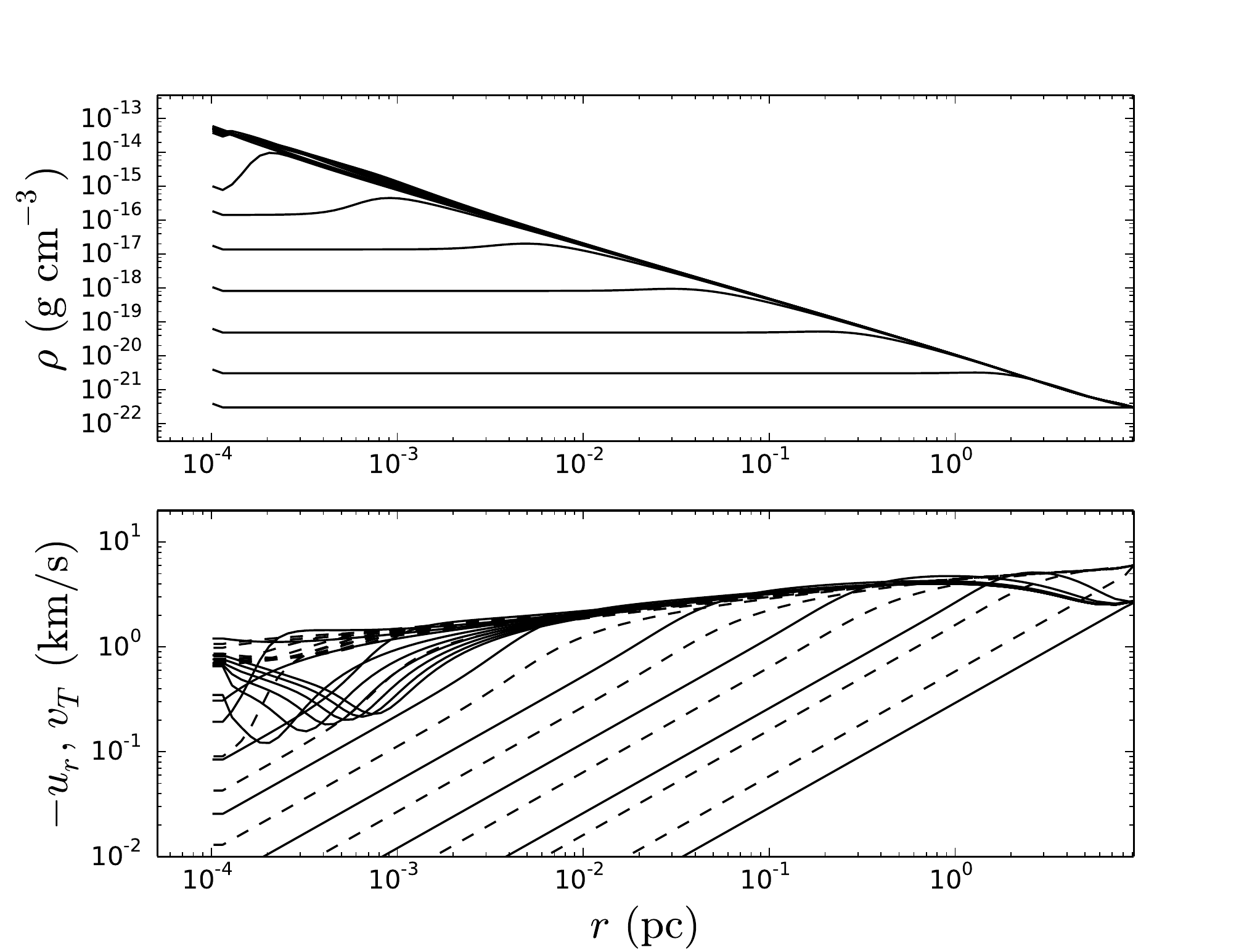}
\caption{\label{fig_rho_v_vs_r_080}The run of density $\rho(r)$ (upper
  panel), infall velocity $-u_r(r)$ (lower panel, solid line), and
  turbulent velocity $v_T(r)$ (lower panel, dashed line) shown for 15
  different times, ranging from the initial conditions to shortly
  after the formation of a star, which occurs at
   $t_*/\tdyn\approx0.253$. The density increases from its initial value
  $\bar\rho=3\times10^{-22}\g\cm^{-3}$ to a maximum
  $\rho\approx10^{-13}\g\cm^{-3}$ at the inner edge of the integration
  region. Starting from the time just before the star forms, the
  density distribution is very nearly independent of time. }
\end{figure}

Figure \ref{fig_rho_v_vs_r_080} shows the run of density and infall
velocity for fifteen snapshots between $t=0\,\tau_{\rm DYN}$ and
$t\approx 0.253\,\tau_{\rm DYN}$, where $\tau_{\rm DYN}\equiv |R/u_r(R)|$. The
first four snapshots show a core in the density distribution,
$\rho(r)=const.$, with the constant increasing rapidly from
$3\times10^{-22}\g\cm^{-3}$ in the initial snapshot to
$10^{-16}\g\cm^{-3}$ in the sixth snapshot. By the eighth snapshot, the
density distribution has evolved into a (slightly broken) power
law. The break in the power law occurs at $r=r_*(t)$; for $r<r_*(t)$,
$\rho(r)\sim r^{1.5}$, while for $r>r_*(t)$, $\rho(r)\sim r^{k_\rho}$,
with $k_\rho\approx1.7$.

The infall velocity $-u_r(r,t)$ also evolves into a broken power
law. For $r<r_*(t)$, the exponent $p'=-0.5$, while for $r>r_*(t)$,
$p'\approx 0.2$, before flattening at $r \approx R$.

\begin{figure}
\epsscale{1.2}
\plotone{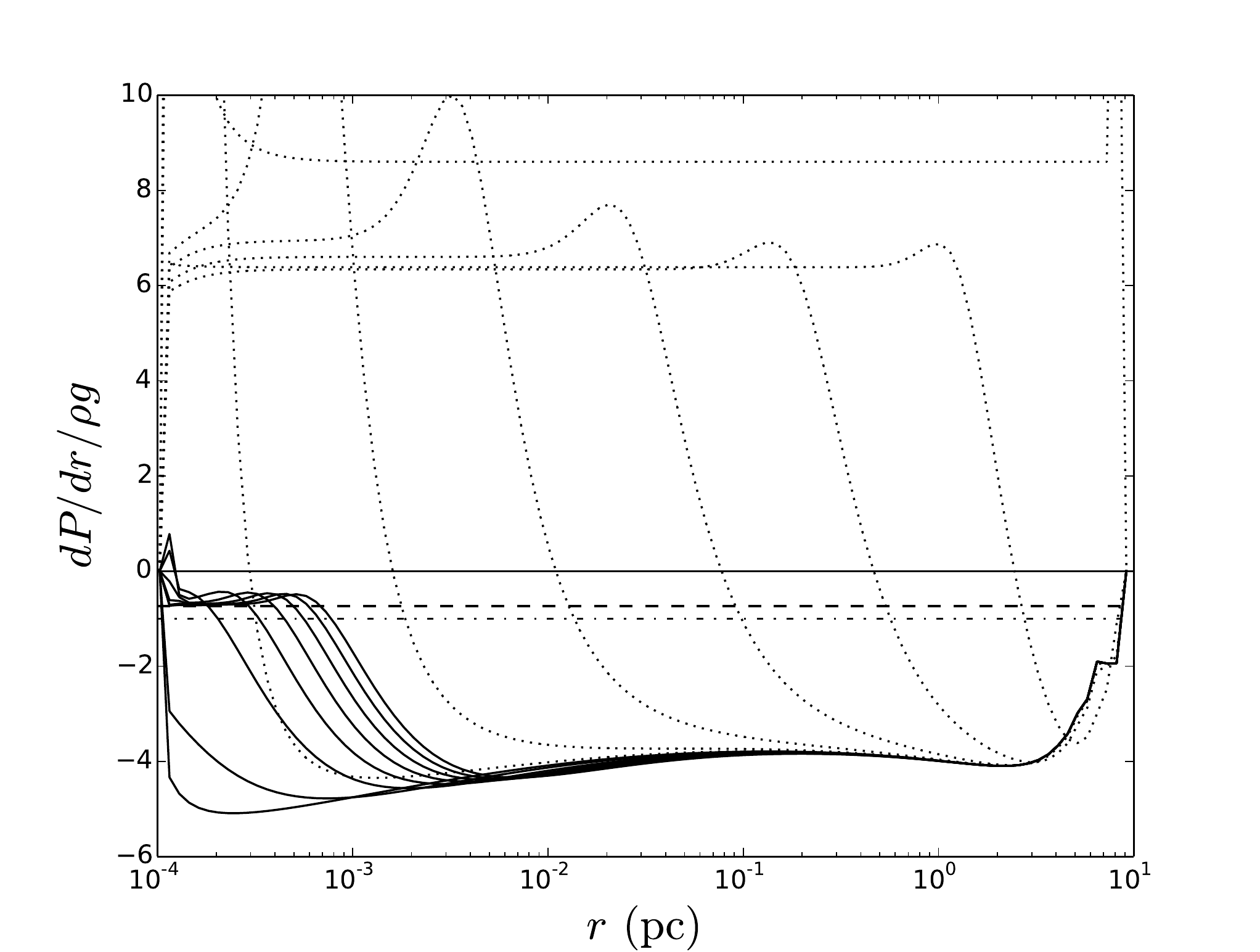}
\caption{\label{fig_pressure_gravity_vs_r}The ratio $(\partial \rho
  v_T^2/\partial r)/(\rho GM(r,t)/r^2)$ plotted versus radius. The
  dotted (solid) lines correspond to times before (after) the central
  star forms. At large $r$ the outward acceleration due to the
  pressure gradient exceeds the inward acceleration due to gravity
  (the dotted and solid lines are below the dot-dashed horizontal line
  at -1); this is why the infall velocity decreases inward for
  $r>r_*(t)$, as seen in figure \ref{fig_rho_v_vs_r_080}. Inside
  $r_*(t)$, the acceleration due to the mass of the star exerts a
  gravitational force on the surrounding gas, but the star provides no
  outward pressure gradient; the small amount of gas between
  $r=r_{in}$ and $r_*$ cannot provide a sufficiently strong pressure
  gradient, and so the magnitude of the infall velocity increases
  inward. However, even after the star forms, the acceleration
  provided by the pressure gradient is always a substantial fraction
  of the total gravitational acceleration---the solid lines are only
  slightly above the dot-dashed line for $r<r_*$. The horizontal
  dashed line is the prediction of Equation (\ref{eqn: small r
    pressure}), at $-5/(5+4\etaeff^2)$, where in this case
  $\etaeff=2/3$. For $r>r_*$, the ratio is roughly consistent with the
  prediction of Equation (\ref{eqn: pressure gradient ratio}).}
\end{figure}

Figure \ref{fig_pressure_gravity_vs_r} shows the ratio
\be  
{(\nabla\rho v_T^2)/\rho \over {GM(r,t)/r^2}}
\ee  
of the pressure gradient term to the gravity term in the momentum
equation. We distinguish between times previous to the formation of
the central object (shown by dotted lines) and times after the central
object forms (shown by solid lines). These curves correspond to the same
times as the curves in the previous figure.

At early times and small radii, the density is nearly constant, so
$(1/\rho)dP/dr\sim dv_T^2/dr>0$, and the pressure gradient provides an
inward force, i.e,., it acts in the same sense as gravity. Thus the
pressure gradient force and the force due to gravity both act to move
gas toward the origin. However, at large radii, the dotted curves drop
down, eventually becoming negative at large radii, i.e.,
the pressure gradient term is directed outward, opposing the force due
to gravity. Near the outer boundary, this outward pressure gradient
term is larger in magnitude than the acceleration of gravity.

After the first time step, the radius at which the pressure gradient
changes sign is located at $r\approx 2\pc$; well inside this radius,
the density is uniform, $\rho(r)=const.$ At each subsequent plotted
time, the transition radius decreases by a factor of about three. Once
again, at a fixed time, but well inside this transition radius, the
density is uniform.

After the central object forms the situation is very different. To
begin with, the pressure gradient is directed outward at all radii.
However while the pressure gradient force is directed outward, it is
smaller in magnitude than the force due to gravity at radii
$r<r_*(t)$. Thus the infall velocity of a (Lagrangian) mass shell
increases inward. The ratio of the two forces given by Equation
(\ref{eqn: small r pressure}) is constant, and is depicted in Figure 2
by the horizontal dashed line. The numerical result at late times
(after the star forms) tracks this prediction for $r<r_*$. As a
result, the infall velocity $u_r\sim r^{-1/2}$, but it is smaller than
the free-fall velocity by a factor $\Gamma/\sqrt{2}\approx 1/2$. This
is different than the prediction of the turbulent core model, in which
the pressure term becomes negligible as $r\to0$, so that the infall
velocity approaches the free-fall value.

In contrast, for $r>r_*(t)$ the pressure gradient term is larger in
magnitude than the gravity term, so that the infall velocity of a mass
shell {\em decreases} inward.  Note that the inertial term
$u_r\partial u_r/\partial r$ is much smaller than either the pressure
gradient term or the gravity term, or their difference, so that the
infall velocity decreases in an Eulerian sense as well, as seen in
Figure \ref{fig_rho_v_vs_r_080}.

\begin{figure}
\epsscale{1.2}
\plotone{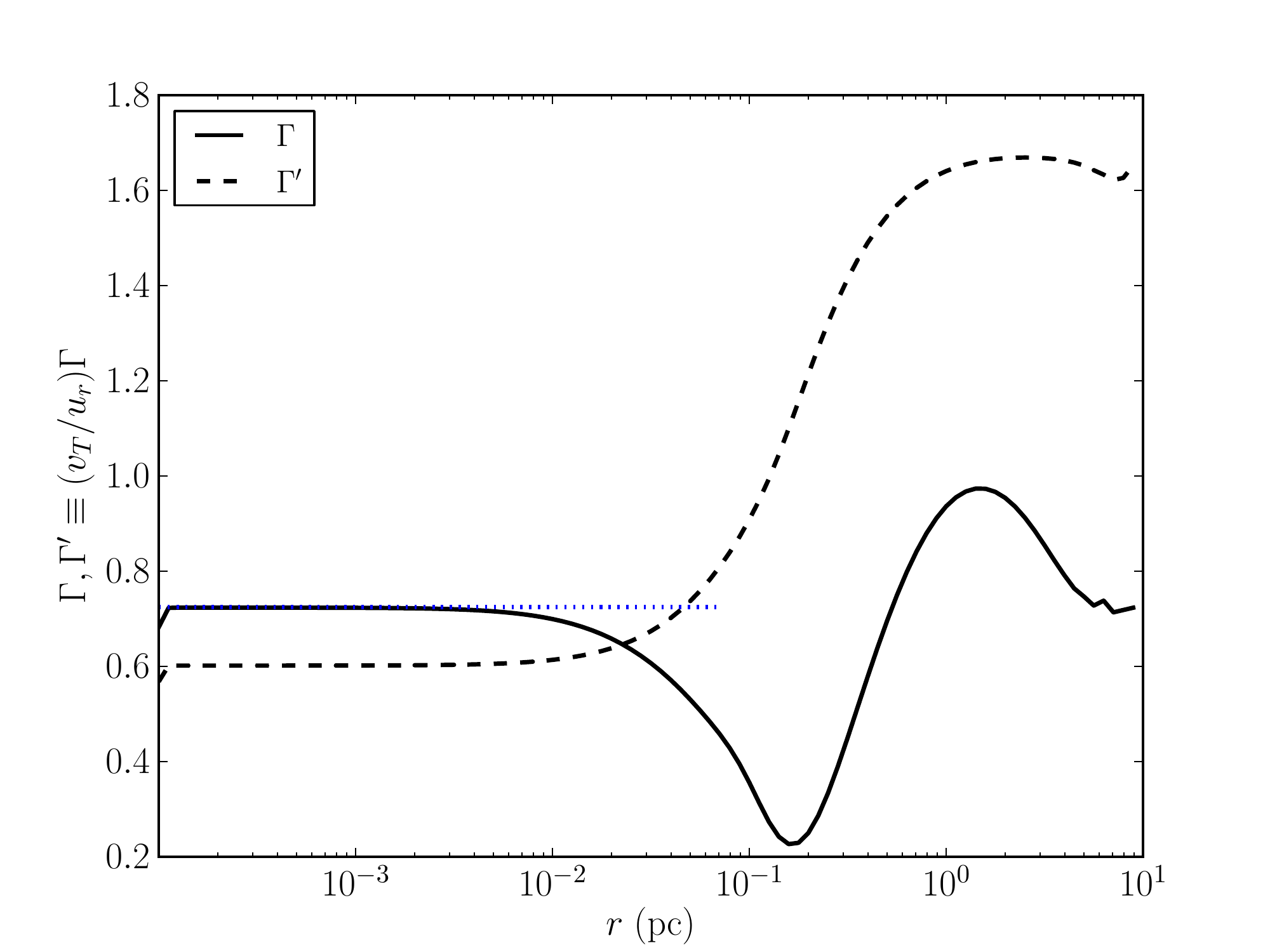}
\caption{\label{fig_Gamma_vs_r_150}The run of $\Gamma(r)$ (solid
  line), and $\Gamma'\equiv (v_T/|u_r|)\Gamma(r)$ (dashed line). The prediction of
  equation (\ref{eqn: Gamma}) at $r=0$,
  $\Gamma=\sqrt{8\etaeff^2/(5+4\etaeff^2)}$, for $r<r_*$ is shown as
  the horizontal dotted line.}
\end{figure}

Figure (\ref{fig_Gamma_vs_r_150}) shows $\Gamma$ and $\Gamma'$. Note
that $\Gamma$ and $\Gamma'$ are nearly constant for $r<r_*$, while $\Gamma'$ is
roughly constant for $r>r_*$. The prediction of Equation (\ref{eqn:
  Gamma}) using the values for the exponents at
$r=0$ is shown in the figure as the horizontal dotted line; it agrees
well with the numerical results.

\begin{figure}
\epsscale{1.2}
\plotone{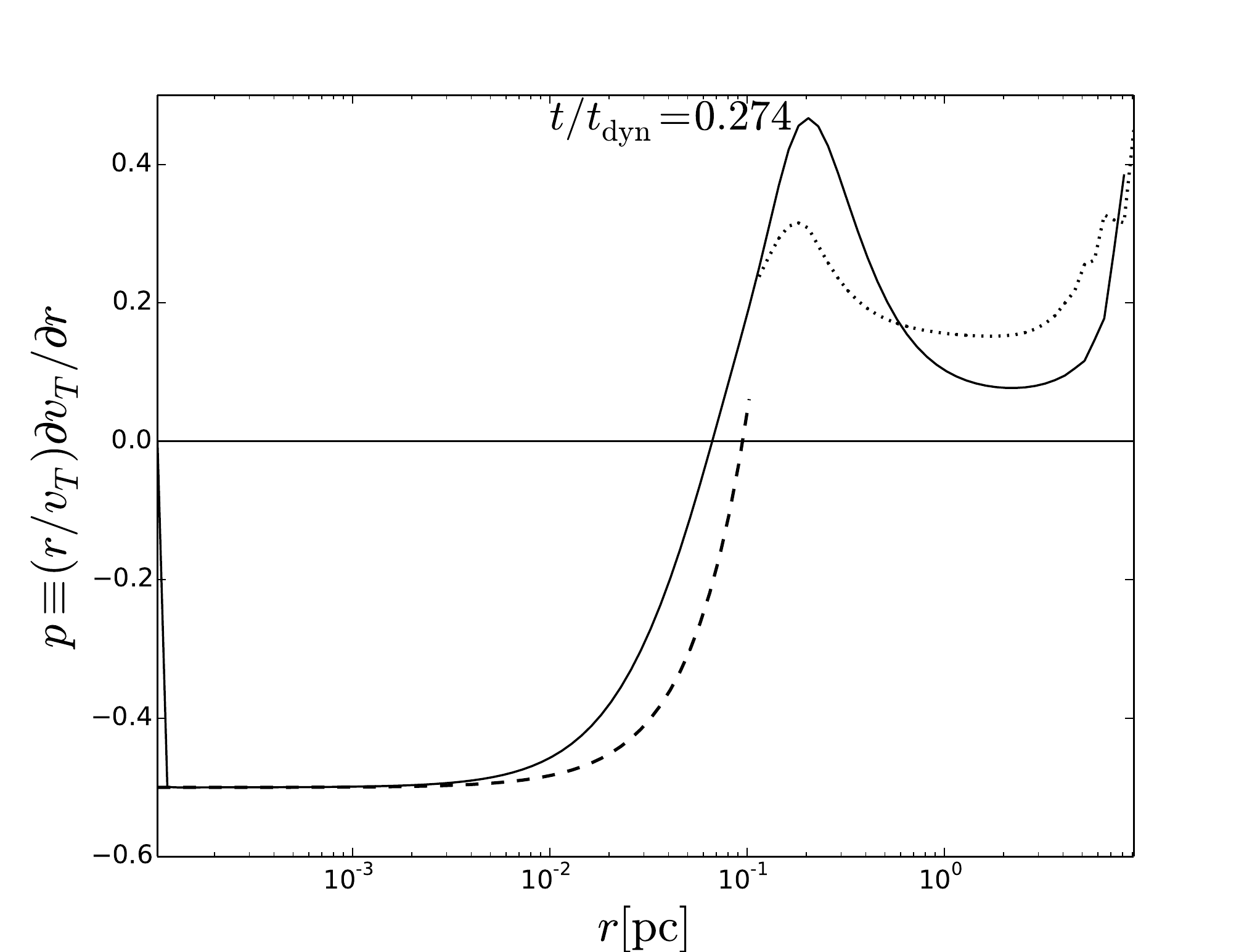}
\caption{\label{fig_p_vs_r_101}The run of $p(r)$, the exponent in the
  size-linewidth relation (solid line),
  and the prediction of Equation (\ref{eqn: p small r}) for $r<r_*$ (dashed line),
  and of Equation (\ref{eqn: p for r big}) for $r_*<r$ (dotted line). }
\end{figure}

Figure (\ref{fig_p_vs_r_101}) shows the exponent $p$ in the
size-linewidth relation, $v_T(r)\sim r^p$. The exponent is positive
for $r>r_*$, but much smaller than the value $p=1/2$ seen in
simulations of non-self-gravitating turbulence. For $r<r_*$ the value
of $p$ quickly plunges to $-0.5$, the value of a Keplerian
profile. However, the normalization is below that of the Keplerian
free-fall velocity as we discussed above. The predictions of the
analytic theory, equations (\ref{eqn: p small r}) for $r<r_*$ and
(\ref{eqn: p for r big}) for $r>r_*$ are also shown, and agree well
with the numerical results except in the immediate vicinity of $r_*$.

\begin{figure}
\epsscale{1.2}
\plotone{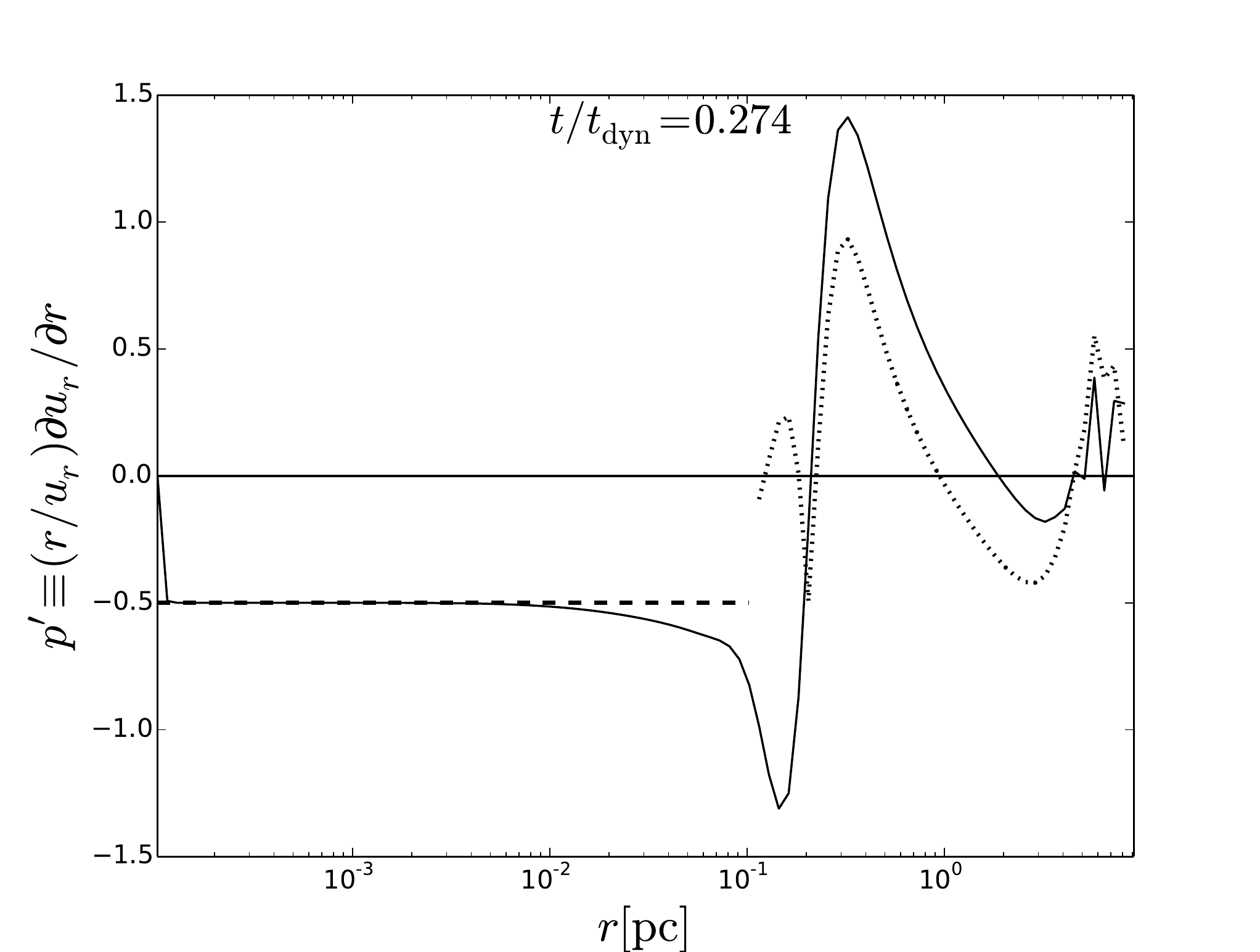}
\caption{\label{fig_p_prime_vs_r_101}The run of $p'(r)$, where $u_r(r)=u_r(R)(r/R)^{p\prime}$
  (solid line),
  and the prediction $p'=-1/2$ for $r<r_*$ (dashed line),
  and of Equation (\ref{eqn: large r p prime}) for $r_*<r$ (dotted line). }
\end{figure}

Figure (\ref{fig_p_prime_vs_r_101}) shows the result for $p'$; the
prediction $p'=-1/2$ for $r<r_*$ and the prediction of Equation
(\ref{eqn: large r p prime}) are shown as the dashed and dotted lines,
respectively.  Here we note again that the gravity from the central
star drives the radial velocity to a Keplerian profile inside $r<r_*$;
while we do not show it here, the magnitude of the infall velocity
does not reach free-fall values.

\begin{figure}
\epsscale{1.2}
\plotone{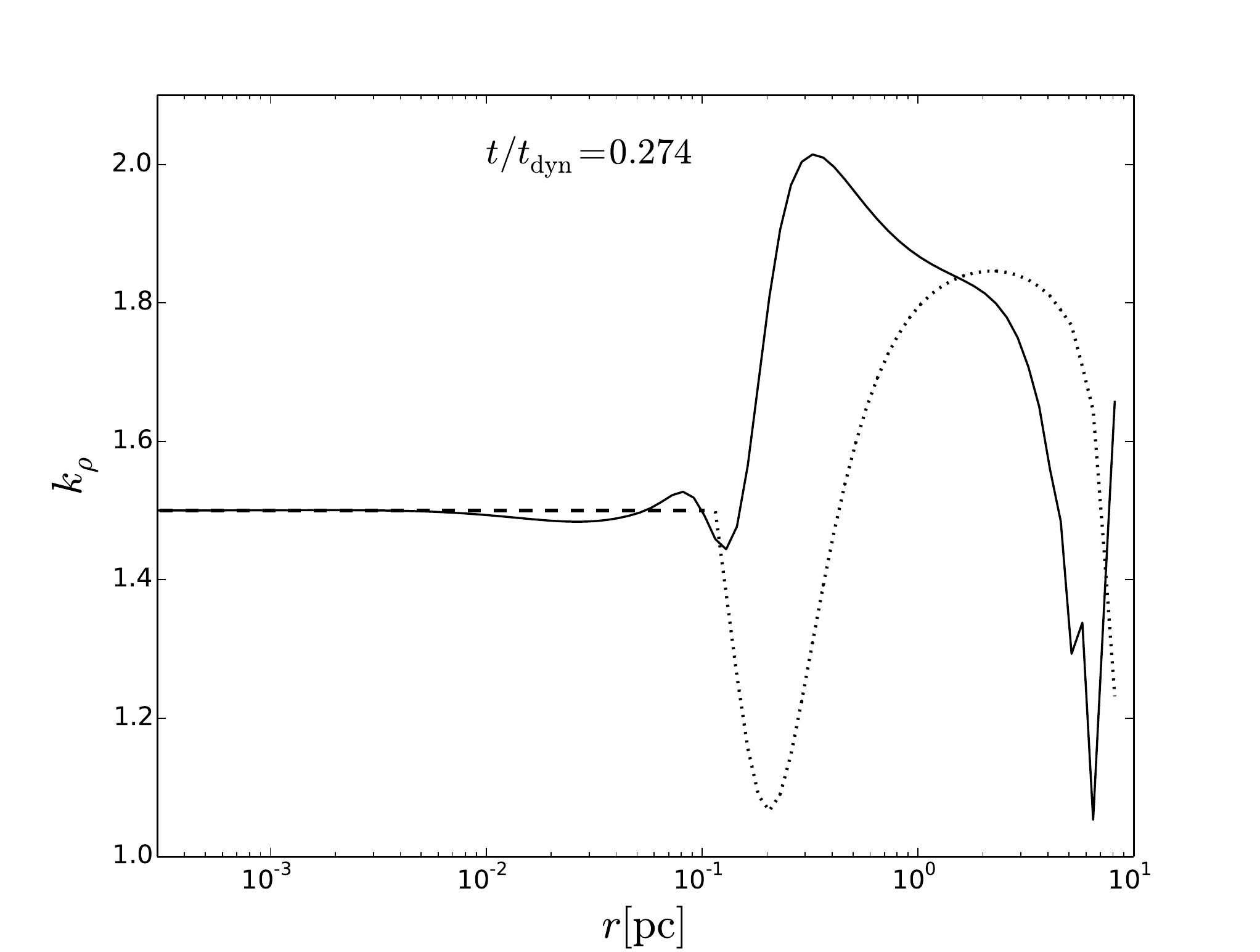}
\caption{\label{fig_k_rho_vs_r_101}The run of $k_\rho(r)$, where $\rho(r)=\rho(R)(r/R)^{-k_\rho}$
  (solid line),
  and the prediction of Equation (\ref{eqn: vanishing r predictions}) for $r<r_*$ (dashed line),
  and of Equation (\ref{eqn: large r momentum}) for $r_*<r$ (dotted line). }
\end{figure}

Figure (\ref{fig_k_rho_vs_r_101}) plots the density exponent $k_\rho$
(solid line) and the prediction of Equation (\ref{eqn: vanishing r
  predictions}) for $r<r_*$ (dashed line), and of Equation (\ref{eqn:
  large r momentum}) for $r_*<r$ (dotted line). The prediction of
$k_{\rho} = 1.5$ is recovered nicely in the numerical results.

\begin{figure}
\epsscale{1.2}
\plotone{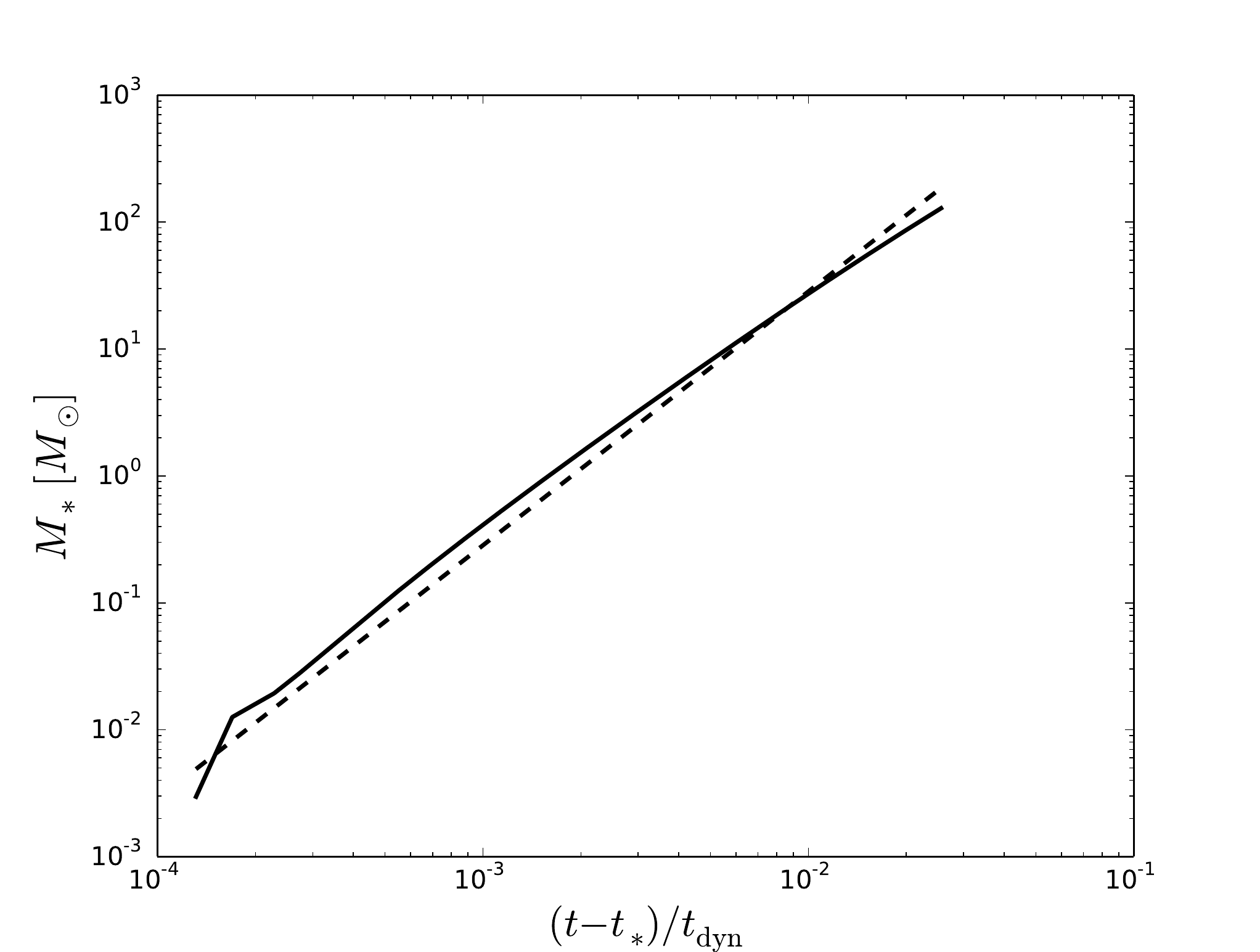}
\caption{\label{fig_mass_vs_time_April_12} The accreted mass as a
  function of time since the star forms, $t-t_*$, where
  $t_*/\tdyn\approx0.25$. The solid line is the numerical result, the
  dashed line is the prediction of Equation (\ref{eqn: mass vs
    time}). }
\end{figure}

Figure (\ref{fig_mass_vs_time_April_12}) shows the accreted mass
versus $t-t_*$, with $t_*$ defined to be the time when the accreted
mass exceeds some minimum mass $M_{\rm min}$; in the figure, $M_{\rm
  min}=10^{-5}M_\odot$. Varying $M_{\rm min}$ alters the appearance of
the plot at early times but has no effect at later times. The mass
increases in proportion to $(t-t_*)^2$, and the numerical result is in
good agreement with the prediction of Equation
(\ref{eqn: mass vs time}).

We have calculated the run of density and velocity when
the effects of a strong magnetic field, corresponding to a plasma
$\beta=1$ relative to $\rho v_T^2$, are
included in the manner described above; in the notation of
Equation (\ref{eqn: magnetic pressure}), $\phi_B=2$. The extra pressure provided by
the magnetic field slows the infall, reducing the rate of accretion
compared to a similar run with no magnetic pressure support by a
factor of 1.9.


\section{DISCUSSION}
\label{sec:discussion}

\subsection{The Connection Between Self-Gravity and the
Size-Linewidth Relation}

Larson's first law, the size-linewidth relation, applies to GMCs as
well as to low mass star forming clumps. The power law index in GMC's
($v_T(r)\sim r^p$) is almost
universally observed to be $p\sim 0.4-0.5$, e.g.,
\citet{1981MNRAS.194..809L,1987ApJ...319..730S,2004ApJ...615L..45H}.
 The scaling relation extends to lower mass and smaller objects, e.g.,
\citet{1983ApJ...270..105M}, who finds $p\approx0.5$.  

In contrast, massive star forming regions show $p\approx 0.2$, and
have much higher levels of turbulence at a given radius than is seen
in low mass star formation regions, e.g., \citet{1992ApJ...384..523F,
  1995ApJ...446..665C,1997ApJ...476..730P}. The turbulent velocities
in these clumps are also much larger than the extrapolation of the GMC
size-linewidth relation predicts. For example
\citet{1997ApJ...476..730P} examine $r\sim0.1-1\pc$ clumps with masses
$\sim 10^3-10^4M_\odot$ selected on the basis of H$_2$O maser
emission. They find turbulent velocities $v_T\approx 2-10\kms$ (FWHM),
nearly a factor of ten above Larson's relation for $r\approx
0.3\pc$---see their Figure 5.

We have shown quantitatively how the
conversion of gravitational potential energy into turbulent motion via
adiabatic heating, coupled with the back pressure produced by that
turbulence, can explain $p\approx0.2$, for $r>r_*$, a result
illustrated in Figure \ref{fig_p_vs_r_101}.

We can extract another prediction: in our model of converging flows,
and in the simulations of Paper I, the density
increases monotonically as $r$ decreases. However, the turbulent
velocity does not vary monotonically with $r$: at large radii
($r>r_*$) the turbulent velocity decreases as $r$ decreases, but
inside the critical radius, the turbulent velocity {\em increases}
with decreasing $r$. It follows that a plot of turbulent velocity
versus density for a single region will show that at low density the
velocity decreases with increasing density, while at high density the
velocity will increase with increasing density. In regions with
different total mass, this breakpoint will occur at different
radii. Even in regions with the same total mass, the break occurs at
different radii, depending on the age of the system, since the
critical radius depends on the mass of the central star or star cluster.

\citet{1997ApJ...476..730P} examine massive clusters and find that the
linewidth increases with density. The gas densities they examine are
very high, of order $n=10^5$ to $10^7\cm^{-3}$. Their highest
luminosity systems exceed $10^6L_\odot$, which require stellar masses
of order $3\times10^3M_\odot$. They give a virial estimate of the
total mass, finding values of order $5000M_\odot$. This suggests that
they may be resolving $r_*$ in their most luminous objects.

Collapse models such as that of
\cite{1997ApJ...476..750M,2003ApJ...585..850M,2004ApJ...615..813F}
predict $u_r(r)$,
but they do not predict the run of $v_T(r)$ at any radius---they take
it as an input. They assume that $v_T(r)\sim r^p$, with $p>0$. Thus
the linewidth will always decrease with increasing density, a
corollary of the model that is inconsistent with the observations of
\citet{1997ApJ...476..730P}, a point those authors noted.

We interpret the finding that $p\approx0.5$ in GMCs and low mass star
formation regions to mean that GMCs, as a whole, are not undergoing
rapid gravitational collapse. This is not to say that the virial
parameter $\alpha_{\rm vir}$ is larger than one; in the most massive
GMCs, $\alpha_{\rm vir}\sim1$. Conversely, we interpret the
observation that $p\approx0.2-0.3$ in compact massive clumps to mean
that they are are the centers of large scale converging flows and are
undergoing rapid gravitational collapse. We note that in the turbulent
collapse model describe here the infall velocity should be
substantially below the free-fall velocity, a point we return to
below.

Papers reporting on simulations of turbulently stirred gas, in which
the turbulence is maintained at a level corresponding to $\alpha_{\rm
  vir}\approx1$, often report the slope of the velocity power
spectrum $P({\bf k})\equiv|v({\bf k})|^2\sim k^{-n_{3D}}$. This slope
is related to the index $p$ by the relation $p=(n_{3D}-3)/2$, or if
the one-dimensional power spectrum is used, $n_{1D}=n_{3D}-2$,
$p=(n_{1D}-1)/2$.  The implied values of $p=0.45$ to $0.5$ in
hydrodynamic turbulence and $p=0.45-0.6$ in MHD simulations
\citep{2007ApJ...661..972P,2006ApJ...638L..25K,2012ApJ...750...13C};
these simulations do not include gravity. However, even simulations
which do include gravity find large values of $n_{1D}$, e.g.,
\citet{2012ApJ...750...13C}, and by implication, large values of $p$.

Paper I (which included the effects of gravity) looked in the vicinity
of local density maxima, which are undergoing collapse, and showed by
direct measurement that $p\approx 0.2-0.3$, substantially smaller than
the global average value in the same turbulent box. That paper also
showed that the velocity power spectrum calculated in the immediate
vicinity of a collapsing region has more power on small scales than
the power spectrum calculated in exactly the same manner around random
points in the same simulation box. The power spectrum, when calculated
over an entire simulation box, is too crude a tool to detect the
presence of collapsing flows.

\subsection{Slow, Radially Extended Infall}\label{sec: slow infall}
The large turbulent pressure support which results from adiabatic
heating tends to reduce the magnitude of the infall velocity, so that
$|u_r(r)|\approx v_T(r)\approx-\Gamma\sqrt{GM_*/r}$ at small radii,
and $|u_r(r)|\lesssim v_T(r)\approx \sqrt{GM(r)/r}\sim r^{+0.2}$ at
large radii. The 3D simulations in Paper I show find a similar result
(see Figure 9 in that paper). The slow increase of the turbulent
velocity with increasing $r$ (for $r>r_*$) is consistent with a
modified Larson's first law, $p\approx0.2$.

\begin{figure}
\epsscale{1.2}
\plotone{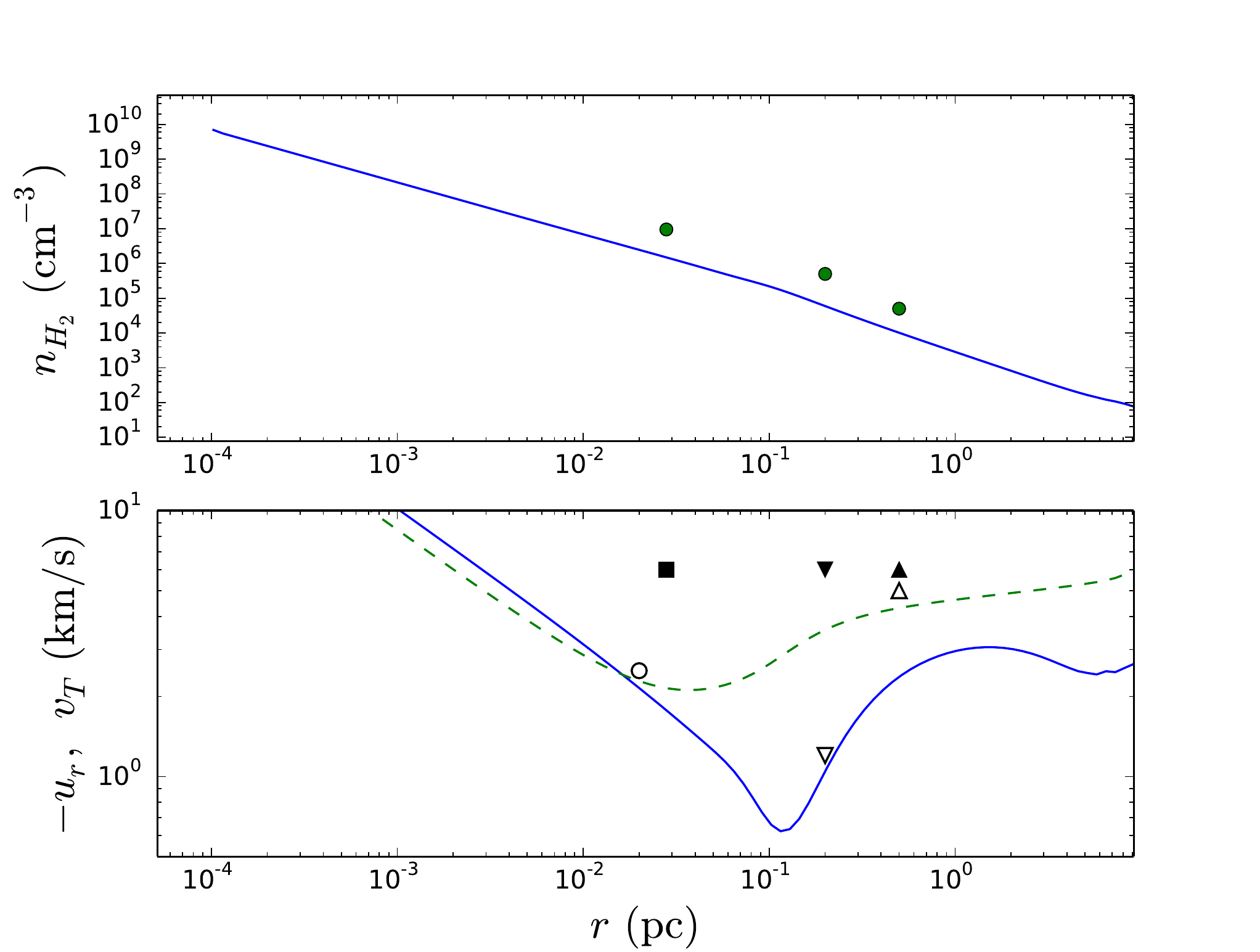}
\caption{\label{fig_rho_v_data} The density, infall, and turbulent
  velocity as a function of radius. The lines give the run of density
  (top panel), and of turbulent and infall velocity (dashed and solid
  lines respectively, in the bottom panel) in the run depicted in
  Figure 1, but at the time when the central mass $M_*=89M_\odot$. The
  points in the upper panel are estimates of the density in the
  galactic massive star forming region G10.62-0.38. In order of
  increasing radius, the data is from \citet{2011A&A...525A.151B},
  \citet{2008ApJ...684.1273K}, and \citet{1986ApJ...304..501H}. The
  points in the lower panel are measurements of $v_T$ (filled symbols)
  and $-u_r$ (open symbols). In order of increasing radius, the
  measured value of $-u_r$ is represented by an open circle
  \citep{2011A&A...530A..53K}, a downward pointing open triangle
  \citep{2008ApJ...684.1273K}, and an upward pointing open triangle
  \citep{1986ApJ...304..501H} The filled square, downward filled
  triangle, and upward filled triangle are from
  \citet{2011A&A...525A.151B}, \citet{2008ApJ...684.1273K}, and
  \citet{1986ApJ...304..501H}; the observational papers do not provide
  an estimate of the errors in the turbulent velocity. The apparent
  dip in $-u_r$ around $r=0.2\pc$ is qualitatively
  consistent with the adiabatically heated turbulent collapse model.}
\end{figure}

These findings contrasts with the expanding collapse wave and related
models, which predict that $v_T$, and hence the pressure, becomes
negligible at small radii, and that the infall velocity reaches the
free-fall value $u_r=-\sqrt{2GM_*/r}$. In fact this class of
models predicts that $|u_r|$ decreases with increasing r at all radii,
or, for expanding collapse wave models, falls to zero at a finite
radius and remains there.  The turbulent velocity is
assumed to be an increasing function of $r$ at all radii.

The different predictions of the infall models should allow us to
distinguish between them. For example, \citet{2008ApJ...684.1273K} see
clear evidence of infall in line profiles of HCO$^+$ at $r=0.3\pc$ in
G10.6-0.4, with $u_r=-1.0\pm0.1\kms$; see Figure
\ref{fig_rho_v_data}. The infall extends inward to $0.03\pc$ as shown
by \citet{2011A&A...530A..53K}, who see inverse P Cygni profiles in CO
lines and infer $u_r=-2.5\kms$. \citet{1986ApJ...304..501H} provide
evidence for infall in the form of absorption in NH$_3$ at
$r\approx0.5\pc$, finding $v_T\approx6\kms$ and $-u_r\approx 5\kms$.

\citet{2011A&A...525A.151B} estimate $v_T=6\kms$ at
$r=0.04(D/6\kpc)\pc$, and a rotational velocity $v_c=2.1\kms$. Hence
in this core, the turbulent pressure dominates both the rotational
energy and the kinetic energy of the infall, by a factor of 9 or
more. Using the masses estimated below, it follows that the infall
velocity is well below the free-fall velocity, by a factor of about
two or three. This is evidence that something is slowing the infall. The fact
that the turbulent velocity is substantially larger than the infall
velocity suggests that turbulent pressure is responsible.

The bolometric luminosity of G10.6-0.4 is $L\approx10^6L_\odot$, and
the stellar mass is therefor at least $M_*\approx 200M_\odot$
\citep{2011ApJ...729..100L}; if the stellar mass function in the
proto-cluster is like the global initial mass function, the stellar
mass could be a thousand solar masses. The gas mass on the scale of
$1.5''\approx 0.04\pc$ is $M_g\sim150M_\odot$
\citep{2009ApJ...703.1308K}. The rough equality of stellar and gas
masses suggests that the observations are probing sizes around
$r_*$. This may explain the increase in the velocity inferred from
$-1\kms$ at $r\approx0.3\pc$ to $-2.5\kms$ at $0.03\pc$, while at the
same time $|u_r|<\sqrt{2GM/r}$. The lines in Figure
\ref{fig_rho_v_data} come from the same model show in Figure
\ref{fig_rho_v_vs_r_080}, but at a later time, specifically, when
$M_*\approx90M_\odot$.

It is worth remarking that the innermost two data points in Figure
\ref{fig_rho_v_data} illustrate a similar phenomenon to that found in
\citet{1997ApJ...476..730P}; the density increases inward, while the
turbulent velocity is flat or possibly slightly increasing inward.

We note that this observation of infall over a decade in radius, from
$r\approx 0.03\pc$ to $0.3\pc$, and possibly to $0.5\pc$, is
consistent with the solutions found in this paper. This observation
does not appear to be consistent with the predictions of the turbulent
core model of \citet{2003ApJ...585..850M}, where the fiducial
hydrostatic core radius $R_{\rm core}=0.05\pc$, and the collapse
proceeds from the inside out, i.e., the infall should be seen only for
$r\lesssim R_{\rm core}$. 

In addition, the apparent increase of $|u_r|$ as $r$ increases beyond
$0.3\pc$ suggested by the observations of \citet{1986ApJ...304..501H}
is inconsistent with collapse models.

\subsection{The Star Formation Efficiency Per Free Fall Time}

We have shown that the star formation rate in self-gravitating
turbulent flows increases linearly in time in the absence of any feedback
physics. This rapid growth in the star formation rate cannot continue
for a dynamical time in real GMCs, since the fraction of gas converted
into stars is seen to be less than $10-20\%$ in the most extreme
cases, and lower in many GMCs. 

We suggest that the low observed star formation efficiency is due to
two types of effects. The first is the effect of early
(pre-supernovae) feedback from the newly formed stars, e.g.,
\citet{2010ApJ...709..191M}.

The second is the effect of initial conditions;
GMCs with large virial parameters will not form stars efficiently,
since they will disperse in less than a GMC free-fall time. This
dispersal cuts off the flow of gas to any collapsing regions in the
cloud, starving the nascent stars. Evidence for this is presented in
paper I, in particular in Figure 14.

Recent papers on the star formation rate in 3D simulations have
generally presented the result in terms of a constant value of $\eff$
\citep[e.g.][]{2011ApJ...730...40P,2012MNRAS.419.3115B,2012ApJ...754...71K}.
The constant $\eff$ that is reported in these papers is found by
fitting a straight line to $M_*(t)$ at late times, typically a
substantial fraction of a free-fall time after the first stars have
formed. However, it is clear from examining the figures in these
papers that $M_*(t)$ is increasing more rapidly than $\sim t$.
Workers that have examined the early time behavior of the star
formation rate in their simulations such as
\citet{2014MNRAS.439.3420M} have found that $\eff$ is not constant
with time.

\subsection{Long-lived Density Structures} 
The theory of adiabatically heated turbulent collapse predicts the
existence of long-lived (compared to the local free-fall time) density
structures. These structures are static in the sense that their shape
does not change; however, they are dynamic in that fluid flows through
them from large (multi-parsec) to small (milli-parsec) scales. For a
significant fraction, of order one quarter, of the large scale
dynamical time, these structures contain no newly formed stars. Seen
in projection, or via extinction, they will have power-law surface
density probability distribution functions. In the early phase, before
the time we have denoted by $t_*$, the gas mass contained in the
structure above a given column will increase with time; after $t_*$ it
will be constant.

Direct observational measurements of $k_\rho$ are difficult, and
exhibit a large range $1.0\lesssim k_\rho\lesssim 2.0$, e.g.,
\citep{2000ApJ...537..283V,2014ApJ...785...42P}, making it difficult
to distinguish between different theoretical models. 

Regions in which $\rho(r)\sim r^{-k_\rho}$ will naturally produce
extinction maps with powerlaw tails to the probability distribution of
extinction \citep{2011ApJ...727L..20K}.  There are only a handful of
observational measurements of the surface density PDF, which should
scale as $\Sigma\sim r^{-2/(k_\rho -1)}$. Power law tails have been
seen in nearby star forming regions, e.g.,
\citet{1999A&A...345..965C,2009ApJ...703...52L,2009A&A...508L..35K}. \citet{2010A&A...512A..67L}
attribute this the effects of gravity. \citet{2011ApJ...727L..20K}
perform numerical simulations of self-gravitating turbulence and find
power-law tails in both the volume density and surface density
distributions. They attribute these tails to collapsing regions. They
also note that if the power law tail of the density $PDF
\sim\rho^{-m}$ is produced by a run of density around density maxima
$\rho(r)\sim r^{-k_\rho}$, then $m=3/k_\rho$.

We agree with this interpretation, with the refinement that the outer
layers of the star forming regions, before self-gravity becomes
dominant, will have a run of density with $k_\rho=1.6-1.8$, while the
gravity-dominated inner region has $k_\rho=1.5$. It is the turbulent
pressure, not the self-gravity, that produces $k_\rho>1.5$, and hence
density PDF tails that have $m<2$.

\subsection{Driving by Protostellar Jets}
Low mass star forming regions show $p=0.5$, similar to that in the ISM
as a whole. Many such regions show prominent proto-stellar jets
\citep{2001ARA&A..39..403R}, which are often suggested to be drivers
of turbulence, e.g., \citet{2005ApJ...632..941Q}. In his analytic
models of turbulence driven by proto-stellar jets,
\citet{2007ApJ...659.1394M} finds $\sigma(r)\sim r^{1/2}$, i.e.,
$p=0.5$ for $r<l$, where $l\approx0.38\pc$ is the characteristic size
of a region subject to protostellar jets with fiducial
properties. Simulations of jet driven turbulence
\citep{2009ApJ...695.1376C} find {\em steeper} spectra,
$p\approx0.87$. The theory does not predict $p\approx0.2$ over an extended
range of radii, whereas the observations show $p\lesssim0.2$ for
$0.04\pc\lesssim r\lesssim 1.6\pc$, e.g., \citep{1995ApJ...446..665C}
or \citep{1997ApJ...476..730P}.

\citet{2004ApJ...615L..45H} note that $p\approx0.5$ over a large range
of scales and for low mass star forming regions, and argue that either
jets produce a feature in $\sigma(r)$ around $r=l$, or the driving due
to outflows is not important even at small scales. In fact, many low
mass star forming regions, e.g., those studied by
\citet{1983ApJ...270..105M}, fall on the original Larson relation,
i.e, with the appropriate magnitude, indicating that neither jets nor
conversion of infall energy to turbulence enhance the turbulence
substantially over that expected from a simple extension of the
behavior on larger scales.

The flat run of turbulent velocity seen in massive star forming
regions, with no indication of a peak on scales of order a parsec,
combined with the considerations above, suggest that the enhanced
turbulence is not due to driving by proto-stellar jets.

\subsection{Local Versus Global Collapse}
The notion that there are coherent collapsing substructures in
turbulent flows in which gravity is important has been suggested in
the recent literature, e.g., \citet{2011MNRAS.411...65B}, and in
particular \citet{2012ApJ...750...13C}. The latter authors stress the
existence of self-similar collapsing spheres at high density. Like
\citet{2011ApJ...727L..20K}, they measured density PDFs $P(\rho)\sim
\rho^m$ with $m\approx1.6-1.8$, and inferred $k_\rho=3/m\approx
1.67-1.88$. It is these spheres, and their dynamics, that we model in
this paper.

\citet{2011MNRAS.411...65B} argue that
  ``...the observational data are consistent with molecular clouds in
  a state of hierarchical and chaotic gravitational collapse,
  i.e. developing local centres of collapse throughout the whole cloud
  while the cloud itself is collapsing, and making equilibrium
  unnecessary at all stages prior to the formation of actual stars.''

We agree that local centers of collapse occur throughout the
cloud. There is direct evidence for local collapse, on parsec scales,
in the form of inverse P Cygni profiles seen toward known star forming
clumps
\citep{2001ApJ...562..770D,2008ApJ...684.1273K,2011A&A...530A..53K,
2011A&A...525A.151B,2013A&A...554A..83V}.

However, we disagree that GMCs (by which we mean clouds with $M\gtrsim
3\times10^5M_\odot$ are collapsing globally. We are not aware of any
detections of inverse P Cygni profiles toward GMCs on scales larger
than $\lesssim10\pc$, while the $10^6M_\odot$ GMCs where the largest
star clusters in the Milky Way form have radii of tens to a hundred
parsecs. We would also point out that the Larson's Law index $p$
measured on large scales is $\sim 0.4-0.5$
\citep{1981MNRAS.194..809L,2011ApJ...740..120R}, substantially larger
than that measured on small scales, typically $p\approx0.1-0.2$
\citep{1995ApJ...446..665C,1997ApJ...476..730P}. If small values of
$p$ are taken as evidence for collapse around star forming regions,
the large values of $p$ seen on large scales must be taken as evidence
against large scale collapse in GMCs. Given enough time, and in the
absence of star formation, GMCs might collapse globally, but in the
Milky Way this does not appear to be happening.

\citet{2007ApJ...657..870V} ran simulations in which two streams of
gas were directed at each other; the subsequent collision generates
turbulence. This is followed, after a dynamical time or two, by a
global gravitational collapse. This collapse drives yet higher
turbulent velocity, with the result that the global kinetic energy is
thereafter similar in magnitude to the global gravitational
energy. This is analogous to our result, but we are examining a local
collapse, whereas theirs is a global result in a simulation
containing a fairly large number of locally collapsing regions
embedded in a globally collapsing cloud.

\citet{2008MNRAS.390..769V} noted that in their simulations, regions
around density maxima showed convergent flows, a result recovered in
the simulations in paper I. They note in their section 6.1 that their
diagnostic tools were not sufficient to decide whether the inflow was
the cause or the consequence of gravitational contraction. We have
shown that it is both---initially convergence leads to an
over-density, and eventually to star formation, whose gravity then
speeds the infall (increases the convergence) on small scales (inside
$r_*$).

\section{CONCLUSIONS}
\label{sec:concl}
We have shown that turbulent self-gravitating flows with virial
parameters near unity have the following properties:

The run of density asymptotes to 
\be
\rho(r,t)=
\begin{dcases}
\rho(r_0)\left({r\over r_0}\right)^{-3/2}, & r<r_*\\
\rho(R,t)\left({r\over R}\right)^{-k_\rho}, \ k_\rho\approx1.6-1.8 & r>r_*.
\end{dcases}
\ee
Since $k_\rho$ is fairly close to $1.5$ at all radii, taking $r_0=R$
is a fair approximation. The prediction that
$\rho(r,t)\to\rho(r)$ for $r<r_*$ is in agreement with the MHD and
HD simulations in Paper I.

This run of density leads to a density PDF of the form $P(\rho)\sim
\rho^{1.6-1.8}$ at high densities, and the related power law for the
surface density PDF. This is in accord both with high resolution three
dimensional MHD and hydro simulations
(e.g. \citealt{2011ApJ...727L..20K}; \citealt{2012ApJ...750...13C};
Paper I), and with observational maps of visual extinction
\citep[e.g.][]{2009ApJ...703...52L,2009A&A...508L..35K}.

The infall velocity 
\be
u_r(r,t)=
\begin{dcases}
-\Gamma\sqrt{GM_*(t)\over r}, \sim r^{-1/2} & r<r_*\\
-\Gamma\sqrt{GM(r,t)\over r} \sim r^{0.2} & r>r_*,
\end{dcases}
\ee
where $\Gamma \approx 0.7$ at small radii, and $\Gamma\approx 1.0$ at
large radii. 

In contrast to most earlier models, the star forming clump is never in
hydrostatic equlibrium. A clear test to decide between the two types
of model is to measure the infall velocity over a large range of
radii; both the expanding collapse wave model and the turbulent core
model predict zero infall velocity at moderate radii (outside the
collapse wave). We noted in \S \ref{sec: slow infall} that there are
some hints that the infall extends to $r\sim0.5\pc$. Models that
select solutions with finite infall velocities at large distance,
e.g., \citet{2004ApJ...615..813F} predict lower values for the infall
velocity at large radii, for a given mass accretion rate and run of
density than the model presented here. In particular, they predict
that $|u_r|$ decreases monotonically as $r$ increases, whereas the
adiabatically heated models predict that $|u_r|$ {\em increases} as
$r$ increases, for $r>r_*$.

The turbulent velocity 
\be
v_T(r,t)=
\begin{dcases}
{1\over 2\etaeff}\Gamma\sqrt{GM_*(t)\over r}, \sim r^{-1/2} & r<r_*\\
{1.2\over \etaeff}\Gamma\sqrt{GM(r,t)\over r} \sim r^{0.2} & r>r_*,
\end{dcases}
\ee
The result $p<0$ for $r<r_*$ explains the observation of
\citet{1997ApJ...476..730P} that the linewidth ($v_T$) increases with
increasing density, a result that they noted was puzzling in the
context of collapse models, since the latter assumed $p>0$ at all radii.

The result that $p\approx0.2$ for $r>r_*$ is the first explanation, to
our knowledge, for the deviation from Larson's law in massive
clumps. In fact, both the larger magnitude of $v_T$ at a given small
distance, and the different slope $p$ are in good agreement with
observations. Both are the direct result of the conversion of infall
kinetic energy and gravitational potential energy to turbulent motions
as the accreting gas falls inward.

The stellar mass increases quadratically with time
\be  
M_*(t)=\phi M_{\rm cl}\left({t-t_*\over \tff}\right)^2.
\ee  

The mass accretion rate 
\be
\dot M(r,t)=
\begin{dcases}
4\pi R^2\rho(R)u_r(r,t), \sim t\,r^{0} & r<r_*\\
4\pi R^2\rho(R)u_r(r,t) \sim t^0\,r^{0.2} & r>r_*.
\end{dcases}
\ee
It may be possible, with ALMA, to measure both the density (using
different tracers) and the infall velocity over a range of radii in
massive star forming clumps.


We would like to thank P. Klaassen for helpful conversations, and for
sharing unpublished results. NM is supported by the Canada Research
Chair program and by NSERC of Canada. PC acknowledges support from the
NASA ATP program through NASA grant NNX13AH43G, and NSF grant
AST-1255469.  This work was supported in part by the National Science
Foundation under Grant No. PHYS-1066293 and the hospitality of the
Aspen Center for Physics. Some of the computations were performed on
the gpc supercomputer at the SciNet HPC Consortium
\citep{2010JPhCS.256a2026L}. SciNet is funded by: the Canada
Foundation for Innovation under the auspices of Compute Canada; the
Government of Ontario; Ontario Research Fund - Research Excellence;
and the University of Toronto. The authors acknowledge the Texas
Advanced Computing Center (TACC) at The University of Texas at Austin
for providing HPC resources that have contributed to the research
results reported within this paper. URL:
\url{http://www.tacc.utexas.edu}

\appendix
\label{sec: appendix}
\section{Estimating the Index $p$ in Larson's Law}
We estimate the power law index $p$ in Larson's Law, for finite
values of $r$, first for $r<r_*$ and then for $r>r_*$. 

\subsection{Estimating $p$ Inside the Sphere of Influence ($r<r_*$)}
We have shown in \S \ref{sec: inside} that $\lim_{r\to0}p=-1/2$. Here
we extend the derivation to finite values, but still for $r<r_*$. 

We start from the momentum Equation (\ref{eqn: momentum}). Using
Equation (\ref{eqn: tau u_r}) and retaining the time derivative,
\be  
{\tdyn\over \tvr}+p'+\left({v_T\over
  u_r}\right)^2(2p-k_\rho)+{GM_*(t)/r\over u_r^2}=0.
\ee  
The last term must vary slowly with $r$, so we find $p'=-1/2$, even at
finite $r$. Since the continuity equation enforces $\partial
\rho/\partial t=0$, Equation (\ref{eqn: rho u}) then ensures that
$k_\rho=3/2$ at finite $r$. However, the time derivative of the infall
velocity is not zero, so $p(r)$ will depend on $r$.

Solving for $p(r)$,
\be  
p(r) = {1\over2}k_\rho -{1\over2}\left({u_r\over  v_T}\right)^2
\left[{\tdyn\over tvr}+p'+\Gamma^{-2}(0)\right],
\ee  
where $\Gamma(0)$ is given by Equation (\ref{eqn: Gamma}). Some algebra
yields
\be \label{eqn: p small r almost} 
p(r)=p(0)-{1\over 2}\left({u_r\over v_T}\right)^2{\tdyn\over \tvr}.
\ee  

To evaluate this, we need an expression for $\tdyn/\tvr$ at small but finite $r$. We start
from Equation (\ref{eqn: u_r small r}), and estimate
\be  \label{eqn: deriv}
{\partial u_r\over\partial t}\approx{1\over 2}u_r {1\over
  M_*(t)}{dM_*(t)\over dt}.
\ee  
Recall that
\be  \label{eqn: mdot appendix} 
{dM_*(t)\over dt}=-\lim_{r\to0}4\pi r^2\rho(r,t)u_r(r,t),
\ee  
while from Equation (\ref{eqn: mass}) we have
%
\be  \label{eqn: mass density} 
4\pi r^2\rho(r,t)=
{3-k_\rho\over r}M_g(r,t).
\ee  
Using eqns. (\ref{eqn: mdot appendix}) and (\ref{eqn: mass density}) in Equation (\ref{eqn: deriv}),
\be  
{\partial u_r\over\partial t}\approx 
-\left({3-k_\rho\over 2}\right){u_r^2\over r}{M_g(r,t)\over M_*(t)}.
\ee  
Thus
\be  \label{eqn: tdyn over tvr}
{\tdyn\over \tvr}\approx
 -\left({3-k_\rho\over 2}\right){M_g(r,t)\over M_*(t)}\approx
-{3\over 4}\left({r\over r_*(t)}\right)^{3/2}
\ee  
for $r<r_*$.

Using this in Equation (\ref{eqn: p small r almost}),
\be \label{eqn: p small r} 
p(r)\approx -{1\over 2} +{3\over 2}\etaeff^2\left({r\over r_*}\right)^{3/2}
\ee  
This prediction is shown in Figure (\ref{fig_p_vs_r_101}) as the dashed line.

\subsection{Estimating $p$ Outside the Sphere of Influence ($r>r_*$)}
The calculation of $p$ for $r<r_*$ was greatly simplified by the fact
that $\partial\rho/\partial t=0$ inside the sphere of influence of the
stars. Outside the sphere of influence, however, we cannot neglect the
time variation of $\rho$, so the continuity equation becomes
\be  \label{eqn: final continuity equation}
{\tdyn\over\trho}+2+p'-k_\rho=0,
\ee  

Combining eqns. (\ref{eqn: final continuity equation}), (\ref{eqn:
  Gprime}), and (\ref{eqn: large r momentum}), we find
\be  \label{eqn: p for r big}
p=\Bigg[
{1\over2}-{1\over (2\Gamma')^2}
-{1\over4}\left({u_r\over v_T}\right)^2
\left(
\left|{\tdyn\over\trho}\right|
+{\tdyn\over\tvt}
\right)
\Bigg ]
\left(1-{1\over2}{u_r^2\over v_T^2}\right)^{-1}.
\ee  
%


To use Equation (\ref{eqn: p for r big}) to evaluate $p$, we need to
estimate the time scale ratios, which depend on the outer boundary
conditions. A simple estimate for the time scale over which the infall
velocity changes is
\be  
{\tdyn\over\tvr}\approx 1.
\ee  
The estimate for the density time scale has to take account of the
outer boundary condition, since in our favored scenario, stars form in
converging flows. At large scales in such flows
$\rho(R)\approx\bar\rho$, where $\bar\rho$ is independent of
time. Thus
\be  
\lim_{r\to R}{\tdyn\over\trho}=0.
\ee  
At smaller radii, but still at radii large compared to $r_*$,
we expect 
$|\tdyn/\rho|\leq 1$.
Since the solution is scale invariant, we estimate
\be  
\left|{\tdyn\over \trho}\right|\approx
\left|{\ln r/r_*\over\ln R/r_*} - 1 \right|.
\ee  

From Equation (\ref{eqn: eliminate}), and anticipating that $p<<1$, we have
\be  
\left({v_T\over u_r}\right)^2\approx \etaeff^2.
\ee  
For our fiducial $\etaeff=2/3$ and $\Gamma'=1.6$, the denominator in Equation (\ref{eqn:
  p for r big}) is roughly $3/4$; as a crude estimate we take
$|\tdyn/\tau_\rho|=1$, so 
\be  
p\approx4/3\left[{1\over 2}-{1\over10}-{1\over4}\cdot{1\over 2}\cdot2\right]\approx0.2
\ee  
as a very rough estimate.

\bibliography{Turbulent}

\end{document}